\documentclass[12pt]{article}
\usepackage{graphics}
\usepackage{graphicx}
\usepackage{color}

\usepackage{amsfonts}

\usepackage{amsmath}
\usepackage{amstext}
\usepackage{amsopn}
\usepackage{amsbsy}
\usepackage{amscd}
\usepackage{amsxtra}
\usepackage{amsthm}
\usepackage{enumerate}
\usepackage{ulem}
\usepackage{psfrag}
\usepackage{lscape}

\makeatletter
\def\hlinewd#1{%
\noalign{\ifnum0=`}\fi\hrule \@height #1 %
\futurelet\reserved@a\@xhline}
\makeatother

\allowdisplaybreaks

\numberwithin{equation}{section}
\numberwithin{figure}{section}
\numberwithin{table}{section}

\renewcommand{\abstractname}

\title{Multi-scale turbulence modeling and maximum information principle. Part 4}

\author{L. Tao\thanks{ Department of Aerospace Engineering, Indian Institute of Technology Madras, Chennai 600 036, India. Email: luoyitao@iitm.ac.in; taoluoyi@gmail.com}}
\date{}

\begin{document}
\maketitle
\def\s{\!}
\def\ss{\!\!}
\def\sss{\!\!\!}
\def\l{\left}
\def\r{\right}
\def\bo{\mathbf{0}}
\def\bx{{\bf x}}
\def\by{{\bf y}}
\def\bz{{\bf z}}
\def\ba{{\bf a}}
\def\bw{{\bf w}}
\def\tbw{\tilde{\bf w}}
\def\bV{{\bf V}}
\def\Vi{V_i}
\def\Vj{V_j}
\def\Vk{V_k}
\def\Vij{\Vi,_j}
\def\Vji{\Vj,_i}
\def\Vkl{\Vk,_l}
\def\bv{{\bf v}}
\def\vi{v_i}
\def\vj{v_j}
\def\tw{\tilde{w}}
\def\wi{w_i}
\def\twi{\tilde{w}_i}
\def\wj{w_j}
\def\twj{\tilde{w}_j}
\def\wk{w_k}
\def\twk{\tilde{w}_k}
\def\wl{w_l}
\def\twl{\tilde{w}_l}
\def\bm{{\bf m}}
\def\bmp{\bm^{\prime}}
\def\bmpp{\bm^{\prime\prime}}
\def\mi{m_i}
\def\bbm{(\bm)}
\def\bbmp{\s\l(\bmp\r)}
\def\bM{{\bf M}}
\def\bbM{\s\l(\bM;\bx\r)}
\def\bn{{\bf n}}
\def\bnp{\bn^{\prime}}
\def\bnpp{\bn^{\prime\prime}}
\def\bbn{(\bn)}
\def\bbnp{\s\l(\bnp\r)}
\def\bk{{\bf k}}
\def\bkp{\bk^{\prime}}
\def\bbk{(\bk)}
\def\bbkp{\s\l(\bkp\r)}
\def\bK{{\bf K}}
\def\bbK{\s\l(\bK;\bx\r)}
\def\bl{{\bf l}}
\def\blp{\bl^{\prime}}
\def\bbl{(\bl)}
\def\bblp{\s\l(\blp\r)}
\def\bi{{\bf i}}
\def\bip{\bi^{\prime}}
\def\bbi{\s\l(\bi\r)}
\def\bbip{\s\l(\bip\r)}
\def\bj{{\bf j}}
\def\bjp{\bj^{\prime}}
\def\bbj{(\bj)}
\def\bbjp{\s\l(\bjp\r)}
\def\bL{{\bf L}}
\def\bp{{\bf p}}
\def\bpp{\bp^{\prime}}
\def\bbp{(\bp)}
\def\bbpp{(\bpp)}
\def\bq{{\bf q}}
\def\bqp{\bq^{\prime}}
\def\p{p}
\def\bbq{(\bq)}
\def\bbqp{\s\l(\bqp\r)}
\def\P{P}
\def\q{q}
\def\tq{\tilde{q}}
\def\barwiwj{\overline{\wi \wj}}
\def\barwiwjwk{\overline{\wi \wj \wk}}
\def\barwixwix{\overline{\wi(\bx)\, \wi(\bx)}}
\def\barwixwjx{\overline{\wi(\bx)\, \wj(\bx)}}
\def\barwixwjy{\overline{\wi(\bx)\, \wj(\by)}}
\def\barwjywkz{\overline{\wj(\by)\, \wk(\bz)}}
\def\barwjywjy{\overline{\wj(\by)\, \wj(\by)}}
\def\barwkzwkz{\overline{\wk(\bz)\, \wk(\bz)}}
\def\barwixwjywkz{\overline{\wi(\bx)\,\wj(\by)\,\wk(\bz)}}
\def\barwixwjxwkx{\overline{\wi(\bx)\,\wj(\bx)\,\wk(\bx)}}
\def\barwipxwjpywkpz{\overline{\wi,_{i'}\ss(\bx)\,\wj,_{j'}\ss(\by)\,\wk,_{k'}\ss(\bz)}}
\def\barwixwjywkzwla{\overline{\wi(\bx,t)\,\wj(\by,t)\,\wk(\bz,t)\,\wl(\ba,t)}}
\def\barbwbw{\overline{\bw\bw}}
\def\bartwimtwjn{\overline{\twi\bbm\twj\bbn}}
\def\bartwimtwln{\overline{\twi\bbm\twl\bbn}}
\def\bartwiktwjl{\overline{\twi\bbk\,\twj\bbl}}
\def\bartwititwjtj{\overline{\twi(t;\bi)\twj(t;\bj)}}
\def\bartwitktwjtl{\overline{\twi(t;\bk)\twj(t;\bl)}}
\def\bartwititwjtjtwktk{\overline{\twi(t;\bi)\,\twj(t;\bj)\,\twk(t;\bk)}}
\def\B{{\cal B}}
\def\DD{{\cal D}}
\def\DDi{\DD\s\l(i\r)}
\def\DDj{\DD\s\l(j\r)}
\def\DDk{\DD\s\l(k\r)}
\def\n{{\rm N}}
\def\nO{{\cal O}\s\l(\n\r)}
\def\b{b}
\def\Aj{A_j}
\def\Al{A_l}
\def\Ak{A_k}
\def\Bi{B_i}
\def\Bj{B_j}
\def\Bk{B_k}
\def\Bl{B_l}
\def\Cji{C_{ji}}
\def\Cli{C_{li}}
\def\Clk{C_{lk}}
\def\Cjk{C_{jk}}
\def\Clm{C_{lm}}
\def\Cnm{C_{nm}}
\def\D{D}
\def\E{E}
\def\EE{{\cal E}}
\def\H{H}
\def\Hij{\H_{ij}}
\def\Wi{W_i}
\def\Wk{W_k}
\def\Wl{W_l}
\def\W{W}
\def\Wij{W_{ij}}
\def\Wil{W_{il}}
\def\Wjl{W_{jl}}
\def\Wkl{W_{kl}}
\def\WiHOMS{\overline{W}_i}
\def\WkHOMS{\overline{W}_k}
\def\WlHOMS{\overline{W}_l}
\def\W{W}
\def\WHOMS{\overline{\W}}
\def\Wij{W_{ij}}
\def\Wil{W_{il}}
\def\Wjl{W_{jl}}
\def\Wkl{W_{kl}}
\def\WijHOMS{\overline{W}_{ij}}
\def\WilHOMS{\overline{W}_{il}}
\def\WjlHOMS{\overline{W}_{jl}}
\def\WklHOMS{\overline{W}_{kl}}
\def\soc{\beta}
\def\socR{\soc^{(R)}}
\def\socI{\soc^{(I)}}
\def\socDim{\soc^{(R)}}
\def\socDimO{\soc^{(a)}}
\def\socS{\Pi}
\def\toc{\gamma}
\def\tocR{\toc^{(R)}}
\def\tocI{\toc^{(I)}}
\def\tocDim{\toc}
\def\tocDimR{\toc^{(R)}}
\def\tocDimI{\toc^{(I)}}
\def\tocDimIO{\toc^{(a)}}
\def\tocDimIOpm{\toc^{(Ia)\pm}}
\def\tocDimIOmp{\toc^{(Ia)\mp}}
\def\tocDimIOp{\toc^{(Ia)+}}
\def\tocDimIOm{\toc^{(Ia)-}}
\def\foc{\delta}
\def\focDim{\hat{\foc}}
\def\foocG{\delta^G}
\def\siocG{\zeta^G}
\def\eiocG{\theta^G}
\def\bN{{\bf N}}
\def\bH{{\bf H}}
\def\bR{\mathbb{R}}
\def\bX{{\bf X}}
\def\barbwbwbw{\overline{\bw\bw\bw}}
\def\G{{\cal G}}
\def\xi{x_i}
\def\xj{x_j}
\def\xk{x_k}
\def\xl{x_l}
\def\balpha{{{\bf \alpha}}}
\def\mbi{b_i}
\def\mbj{b_j}
\def\mbk{b_k}
\def\inform{I}
\def\pdf{f}
\def\pdfG{f_G}
\def\pdfD{f_D}
\def\pdfL{f^{(L)}}
\def\pdfH{f^{(H)}}
\def\varvec{\hat{\hat{\bw}}}
\def\Reprevarvec{\hat{\hat{\bw}}}
\def\ReprevarvecL{\hat{\hat{\bw}}^{(L)}}
\def\ReprevarvecH{\hat{\hat{\bw}}^{(H)}}
\def\bkH{{\bf k}^{(H)}}
\def\LagMultiplier{\lambda}
\def\bK{{\bf K}}
\def\hhw{\hat{\hat{w}}}
\def\Det{\inform_D}
\def\K{\inform_T}
\def\vort{\omega}
\def\BETA{\pmb{\beta}}
\def\ip{i^{\prime}}
\def\jp{j^{\prime}}
\def\kp{k^{\prime}}
\def\lp{l^{\prime}}
\def\twip{\tilde{w}_{\ip}}
\def\twjp{\tilde{w}_{\jp}}
\def\twkp{\tilde{w}_{\kp}}
\def\twlp{\tilde{w}_{\lp}}
\def\imaginary{\imath}
\def\Real{\text{RE}}
\def\Imag{\text{IM}}
\def\VGrad{V}
\def\twl{\tilde{w}_l}
\def\twm{\tilde{w}_m}
\def\twn{\tilde{w}_n}
\def\twmp{\tilde{w}_{m'}}
\def\twnp{\tilde{w}_{n'}}
\def\twmpp{\tilde{w}_{m''}}
\def\twnpp{\tilde{w}_{n''}}
\def\bm{{\bf m}}
\def\bmp{\bm^{\prime}}
\def\bmpp{\bm^{\prime\prime}}
\def\mi{m_i}
\def\bbm{(\bm)}
\def\bbmp{\s\l(\bmp\r)}
\def\bM{{\bf M}}
\def\bbM{\s\l(\bM;\bx\r)}
\def\bn{{\bf n}}
\def\bnp{\bn^{\prime}}
\def\bnpp{\bn^{\prime\prime}}
\def\bbn{(\bn)}
\def\bbnp{\s\l(\bnp\r)}
\def\bk{{\bf k}}
\def\bkp{\bk^{\prime}}
\def\bbk{(\bk)}
\def\bbkp{\s\l(\bkp\r)}
\def\bK{{\bf K}}
\def\bbK{\s\l(\bK;\bx\r)}
\def\bl{{\bf l}}
\def\blp{\bl^{\prime}}
\def\bbl{(\bl)}
\def\bblp{\s\l(\blp\r)}
\def\bi{{\bf i}}
\def\bip{\bi^{\prime}}
\def\bbi{\s\l(\bi\r)}
\def\bbip{\s\l(\bip\r)}
\def\bj{{\bf j}}
\def\bjp{\bj^{\prime}}
\def\bbj{(\bj)}
\def\bbjp{\s\l(\bjp\r)}
\def\bL{{\bf L}}
\def\bp{{\bf p}}
\def\bpp{\bp^{\prime}}
\def\bbp{(\bp)}
\def\bbpp{(\bpp)}
\def\bq{{\bf q}}
\def\bqp{\bq^{\prime}}
\def\p{p}
\def\bbq{(\bq)}
\def\bbqp{\s\l(\bqp\r)}
\def\P{P}
\def\q{q}
\def\tq{\tilde{q}}
\def\barqq{\overline{\q \q}}
\def\barwiwj{\overline{\wi \wj}}
\def\barwiwjwk{\overline{\wi \wj \wk}}
\def\barwiwjwkwl{\overline{\wi \wj \wk \wl}}
\def\barwixwjy{\overline{\wi(\bx)\, \wj(\by)}}
\def\barwkxwkx{\overline{\wk(\bx)\, \wk(\bx)}}
\def\barwixwjywkz{\overline{\wi(\bx)\,\wj(\by)\,\wk(\bz)}}
\def\barwixwjywkzwla{\overline{\wi(\bx)\,\wj(\by)\,\wk(\bz)\,\wl(\ba)}}
\def\barbwbw{\overline{\bw\bw}}
\def\bartwimtwjn{\overline{\twi\bbm\twj\bbn}}
\def\bartwimtwln{\overline{\twi\bbm\twl\bbn}}
\def\bartwiktwjl{\overline{\twi\bbk\,\twj\bbl}}
\def\bartwititwjtj{\overline{\twi(t;\bi)\twj(t;\bj)}}
\def\bartwitktwjtl{\overline{\twi(t;\bk)\twj(t;\bl)}}
\def\bartwititwjtjtwktk{\overline{\twi(t;\bi)\,\twj(t;\bj)\,\twk(t;\bk)}}
\def\bartwiitwjj{\overline{\twi(\bi)\twj(\bj)}}
\def\bartwiktwjl{\overline{\twi(\bk)\twj(\bl)}}
\def\bartwiitwjjtwkk{\overline{\twi(\bi)\,\twj(\bj)\,\twk(\bk)}}
\def\ReynoldsNo{\text{Re}}
\def\tDim{\hat{t}}
\def\WNS{{\cal W}}
\def\ubk{\breve{\mathbf{k}}}
\def\ubm{\breve{\mathbf{m}}}
\def\ubn{\breve{\mathbf{n}}}
\def\ubl{\breve{\mathbf{l}}}
\def\ubj{\breve{\mathbf{j}}}
\def\uk{\breve{k}}
\def\um{\breve{m}}
\def\un{\breve{n}}
\def\ul{\breve{l}}
\def\usoc{\breve{\soc}}
\def\usocDim{\breve{\soc}}
\def\usocDimO{\breve{\soc}^{(a)}}
\def\utocDim{\breve{\toc}}
\def\utocDimR{\breve{\toc}^{(R)}}
\def\utocDimI{\breve{\toc}^{(I)}}
\def\utocDimRO{\breve{\toc}^{(Ra)}}
\def\utocDimIO{\breve{\toc}^{(Ia)}}
\def\transformedsoc{\hat{\soc}}
\def\transformedsocDim{\hat{\soc}}
\def\transformedsocDimO{\hat{\soc}^{(a)}}
\def\transformedtocDim{\hat{\toc}}
\def\transformedtocDimR{\hat{\toc}^{(R)}}
\def\transformedtocDimI{\hat{\toc}^{(I)}}
\def\transformedtocDimRO{\hat{\toc}^{(Ra)}}
\def\transformedtocDimIO{\hat{\toc}^{(Ia)}}
\def\transformedbk{\hat{\mathbf{k}}}
\def\transformedbm{\hat{\mathbf{m}}}
\def\transformedbn{\hat{\mathbf{n}}}
\def\transformedbl{\hat{\mathbf{l}}}
\def\transformedbj{\hat{\mathbf{j}}}
\def\transformedk{\hat{k}}
\def\transformedm{\hat{m}}
\def\transformedn{\hat{n}}
\def\transformedl{\hat{l}}
\def\betaT{\sqrt{\s\s\beta}}
\def\g{\gamma}
\def\gO{\g^{(a)}}
\def\bO{\betaT^{(a)}}
\def\B{B}
\def\G{G}

\def\br{\mathbf{r}}
\def\bs{\mathbf{s}}
\def\wm{w_m}
\def\wn{w_n}
\def\W{U}
\def\Wij{\W_{ij}}
\def\Wkj{\W_{kj}}
\def\Wji{\W_{ji}}
\def\Wikj{\W_{ikj}}
\def\Wkij{\W_{kij}}
\def\Wjki{\W_{jki}}
\def\Wkji{\W_{kji}}
\def\Wjlk{\W_{jlk}}
\def\Wlkj{\W_{lkj}}
\def\Q{Q}
\def\Qi{\Q_i}
\def\Qj{\Q_j}
\def\Qk{\Q_k}
\def\tW{\tilde{U}}
\def\tWij{\tW_{ij}}
\def\tWkj{\tW_{kj}}
\def\tWji{\tW_{ji}}
\def\tWikj{\tW_{ikj}}
\def\tWkij{\tW_{kij}}
\def\tWjki{\tW_{jki}}
\def\tWkji{\tW_{kji}}
\def\tWjlk{\tW_{jlk}}
\def\tWjli{\tW_{jli}}
\def\tWlkj{\tW_{lkj}}
\def\tWlki{\tW_{lki}}
\def\tQ{\tilde{Q}}
\def\tQi{\tQ_i}
\def\tQj{\tQ_j}
\def\tQk{\tQ_k}

\def\tWOneOneDim{\beta^{(R)}}
\def\tWOneOneOneDim{\gamma^{(I)}_{1}}
\def\tWOneTwoOneDim{\gamma^{(I)}_{2}}
\def\ok{\overline{k}}
\def\obk{\overline{\bk}}
\def\obr{\overline{\br}}
\def\ot{\overline{t}}

\def\socAsy{\beta^{(a)}}
\def\tocAsy{\gamma^{(a)}}
\def\focAsy{\delta^{(a)}}
\def\foc{\delta}
\def\fociAsy{\foc^{(a)}_i}
\def\focIAsy{\foc^{(a)}_1}
\def\focIIAsy{\foc^{(a)}_2}
\def\focIIIAsy{\foc^{(a)}_3}
\def\foci{\foc_i}
\def\focI{\foc_1}
\def\focII{\foc_2}
\def\focIII{\foc_3}
\def\QAsy{Q^{(a)}}
\def\tQAsy{\tQ^{(a)}}
\def\WAsy{\W^{(a)}}
\def\tWAsy{\tW^{(a)}}
\def\kp{k^{\prime}}
\def\lp{l^{\prime}}
\def\bkp{\bk^{\prime}}
\def\blp{\bl^{\prime}}
\def\Kinfty{K^{\infty}}
\def\psiAsy{\psi^{(a)}}
\def\tWtocAsy{\tW^{(Ia)}}

\def\gW{\dot{\W}}
\def\gQ{\dot{\Q}}
\def\gsoc{\dot{\soc}}
\def\gtoc{\dot{\toc}}
\def\gfoc{\dot{\foc}}
\def\gfocI{\dot{\focI}}
\def\gfocII{\dot{\focII}}
\def\gsocAsy{\gsoc^{(a)}}
\def\gtocAsy{\gtoc^{(a)}}
\def\gfocAsy{\dot{\foc}^{(a)}}
\def\gfocIAsy{\dot{\foc}^{(a)}_1}
\def\gfocIIAsy{\dot{\foc}^{(a)}_2}
\def\ggfoc{\ddot\foc}

\def\dsoc{\dot{\soc}}
\def\dtoc{\dot{\toc}}
\def\dfoc{\dot{\foc}}
\def\dfocI{\dot{\foc}_1}
\def\dfocII{\dot{\foc}_2}
\def\dsocAsy{\dot{\soc}^{(a)}}
\def\dtocAsy{\dot{\toc}^{(a)}}
\def\dfocAsy{\dot{\foc}^{(a)}}
\def\dfocIAsy{\dot{\foc}^{(a)}_1}
\def\dfocIIAsy{\dot{\foc}^{(a)}_2}

\def\w{w}
\def\tw{\tilde{\w}}
\def\underlinei{\sout{i}}
\def\underlinej{\sout{j}}
\def\underlinek{\sout{k}}
\def\underlinel{\sout{l}}
\def\underlinem{\sout{m}}
\def\underlinen{\sout{n}}
\def\underlineI{\sout{I}}
\def\underlineJ{\sout{J}}
\def\underlineK{\sout{K}}
\def\underlineL{\sout{L}}
\def\underlineM{\sout{M}}
\def\underlineN{\sout{N}}

\def\tk{\tilde{k}}
\def\tbk{\tilde{\bk}}

\def\SS{s}
\def\CC{c}
\def\Kinfty{K^{\infty}}

\def\MPeakTwo{k_{2P}}
\def\MValleyTwo{k_{2V}}
\def\MPeakOne{k_{1\text{min}}}
\def\MValleyOne{k_{1\text{max}}}
\def\MMiddle{k_{20}(k_1;\sigma)}
\def\SupportNbkp{{\cal S}_N(\sigma)}
\def\Supportfbkp{{\cal S}_f(\sigma)}
\def\unitkone{\bar{k}_1}
\def\unitktwo{\bar{k}'_2}
\def\unitlone{\bar{l}_1}
\def\unitltwo{\bar{l}_2}

\def\konemin{k_{1min}(\sigma)}
\def\koneminzero{k_{1min}(0)}
\def\MExtremeDomain{{\cal S}_M(\sigma)}
\def\MExtremeDomainVoid{{\cal S}_M}
\def\MaxSupportFbkpbl{{\cal S}_{I\s I\s I}(\sigma)}
\def\MaxSupportNbkp{{\cal S}_N(\sigma)}
\def\MaxSupportNbkpVoid{{\cal S}_N}
\def\MaxSupportNbkpNeg{{\cal S}_N^{-}(\sigma)}
\def\MaxSupportNbkpNegVoid{{\cal S}_N^{-}}
\def\MaxSupportNbkpPos{{\cal S}_N^{+}(\sigma)}
\def\MaxSupportNbkpPosVoid{{\cal S}_N^{+}}
\def\MaxSupportFbkpblNodes{{\cal N}_{I\s I\s I}(\sigma)}
\def\MaxSupportNbkpNodes{{\cal N}_{N}(\sigma)}
\def\MaxSupportNbkpNegNodes{{\cal N}_{N}^{-}(\sigma)}
\def\MaxSupportNbkpPosNodes{{\cal N}_{N}^{+}(\sigma)}
\def\MaxSupportFbkpblFvalues{{\cal F}(\sigma)}
\def\MaxSupportFbkpblFvaluesNeg{{\cal F}^{-}(\sigma)}
\def\MaxSupportFbkpblFvaluesPos{{\cal F}^{+}(\sigma)}

\def\MPeakTwo{k_{2P}}
\def\MValleyTwo{k_{2V}}
\def\MPeakOne{k_{1\text{min}}}
\def\MValleyOne{k_{1\text{max}}}
\def\MMiddle{k_{20}(k_1;\sigma)}
\def\Supportgammabbbkp{{\cal S}_{\gamma^{(a)}_{22}}(\sigma)}
\def\Supportfbkp{{\cal S}_f(\sigma)}
\def\unitkone{\bar{k}_1}
\def\unitktwo{\bar{k}'_2}
\def\unitlone{\bar{l}_1}
\def\unitltwo{\bar{l}_2}

\def\konemin{k_{1min}(\sigma)}
\def\koneminzero{k_{1min}(0)}
\def\MExtremeDomain{{\cal S}_M(\sigma)}
\def\MExtremeDomainVoid{{\cal S}_M}
\def\MaxSupportFbkpbl{{\cal S}_{I\s I\s I}(\sigma)}
\def\MaxSupportgammabbbkp{{\cal S}_{\gamma^{(a)}_{22}}(\sigma)}
\def\MaxSupportgammabbbkpVoid{{\cal S}_{\gamma^{(a)}_{22}}}
\def\MaxSupportgammabbbkpNeg{{\cal S}_{\gamma^{(a)}_{22}}^{-}(\sigma)}
\def\MaxSupportgammabbbkpNegVoid{{\cal S}_{\gamma^{(a)}_{22}}^{-}}
\def\MaxSupportgammabbbkpPos{{\cal S}_{\gamma^{(a)}_{22}}^{+}(\sigma)}
\def\MaxSupportgammabbbkpPosVoid{{\cal S}_{\gamma^{(a)}_{22}}^{+}}
\def\MaxSupportFbkpblNodes{{\cal N}_{I\s I\s I}(\sigma)}
\def\MaxSupportgammabbbkpNodes{{\cal N}_{\gamma^{(a)}_{22}}(\sigma)}
\def\MaxSupportgammabbbkpNegNodes{{\cal N}_{\gamma^{(a)}_{22}}^{-}(\sigma)}
\def\MaxSupportgammabbbkpPosNodes{{\cal N}_{\gamma^{(a)}_{22}}^{+}(\sigma)}

\def\I{I}
\def\J{J}
\def\M{M}
\def\Incrementione{\hat{\I}_1}
\def\Incrementitwo{\hat{\I}_2}
\def\Incrementjone{\hat{\J}_1}
\def\Incrementjtwo{\hat{\J}_2}
\def\MaxSupportFNbkp{{\cal S}_{N_F}(\sigma)}
\def\MaxSupportFNbkpNeg{{\cal S}_{N_F}^{-}(\sigma)}
\def\MaxSupportFNbkpPos{{\cal S}_{N_F}^{+}(\sigma)}
\def\MaxSupportGbkpbl{{\cal S}_{\gamma}(\sigma)}
\def\MaxSupportGNbkp{{\cal S}_{N_{\gamma}}(\sigma)}
\def\CharacteristicFunction{\chi_{\MaxSupportNbkp}}

\def\LPObjectiveFunctionCoefficient{c}
\def\LPConstraintCoefficient{a}
\def\MaxSupportbeta{{\cal S}_{\beta^{(a)}}(\sigma)}
\def\MaxSupportbetaVoid{{\cal S}_{\beta}}
\def\MaxSupportbetaUB{k_{2U\s B}}

\def\konemin{k_{1min}(\sigma)}
\def\koneminzero{k_{1min}(0)}
\def\MExtremeDomain{{\cal S}_M(\sigma)}
\def\MExtremeDomainVoid{{\cal S}_M}
\def\MaxSupportFbkpbl{{\cal S}_{I\s I\s I}(\sigma)}
\def\MaxSupportNbkp{{\cal S}_N(\sigma)}
\def\MaxSupportNbkpVoid{{\cal S}_N}
\def\MaxSupportNbkpNeg{{\cal S}_N^{-}(\sigma)}
\def\MaxSupportNbkpNegVoid{{\cal S}_N^{-}}
\def\MaxSupportNbkpPos{{\cal S}_N^{+}(\sigma)}
\def\MaxSupportNbkpPosVoid{{\cal S}_N^{+}}
\def\MaxSupportGbkpblNodes{{\cal N}_{\gamma}(\sigma)}
\def\MaxSupportNbkpNodes{{\cal N}_{N}(\sigma)}
\def\MaxSupportNbkpNegNodes{{\cal N}_{N}^{-}(\sigma)}
\def\MaxSupportNbkpPosNodes{{\cal N}_{N}^{+}(\sigma)}
\def\MaxSupportFbkpblFvalues{{\cal F}(\sigma)}
\def\MaxSupportFbkpblFvaluesNeg{{\cal F}^{-}(\sigma)}
\def\MaxSupportFbkpblFvaluesPos{{\cal F}^{+}(\sigma)}

\def\I{I}
\def\J{J}
\def\M{M}
\def\Incrementione{\hat{\I}_1}
\def\Incrementitwo{\hat{\I}_2}
\def\Incrementjone{\hat{\J}_1}
\def\Incrementjtwo{\hat{\J}_2}
\def\MaxSupportFNbkp{{\cal S}_{N_F}(\sigma)}
\def\MaxSupportFNbkpNeg{{\cal S}_{N_F}^{-}(\sigma)}
\def\MaxSupportFNbkpPos{{\cal S}_{N_F}^{+}(\sigma)}
\def\MaxSupportGbkpbl{{\cal S}_{\gamma}(\sigma)}
\def\MaxSupportGNbkp{{\cal S}_{N_{\gamma}}(\sigma)}
\def\CharacteristicFunction{\chi_{\MaxSupportNbkp}}
\def\CharacteristicFunctionNeg{\chi_{\MaxSupportNbkpNeg}}

\def\LPObjectiveFunctionCoefficient{c}
\def\LPConstraintCoefficient{a}
\def\MaxSupportbeta{{\cal S}_{\beta}(\sigma)}
\def\MaxSupportbetaVoid{{\cal S}_{\beta}}

\def\NodeSNNeg{N^{-}}
\def\TriangleSNNeg{T^{-}}
\def\PointMatrixSNNeg{{\cal P}^{-}_N(\sigma)}
\def\PointMatrixSNNegVoid{{\cal P}^{-}_N}
\def\ConnectivityMatrixSNNeg{{\cal C}^{-}_N(\sigma)}
\def\ConnectivityMatrixSNNegVoid{{\cal C}^{-}_N}
\def\NodeNumberTotalSNNeg{N_{}{\raisebox{-1pt}{\text{\tiny $\PointMatrixSNNegVoid$}}}}
\def\TriangleNumberTotalSNNeg{N_{}{\raisebox{-1pt}{\text{\tiny $\ConnectivityMatrixSNNegVoid$}}}} 
\def\MaxSupportNbkpNegTriangles{{\cal T}_{N}^{-}(\sigma)}
\def\MaxSupportNbkpNegTrianglesVoid{{\cal T}_{N}^{-}}
\def\NodeSNPos{N^{+}}
\def\TriangleSNPos{T^{+}}
\def\PointMatrixSNPos{{\cal P}^{+}_N(\sigma)}
\def\PointMatrixSNPosVoid{{\cal P}^{+}_N}
\def\ConnectivityMatrixSNPos{{\cal C}^{+}_N(\sigma)}
\def\ConnectivityMatrixSNPosVoid{{\cal C}^{+}_N}
\def\NodeNumberTotalSNPos{N_{\text{\tiny $\PointMatrixSNPosVoid$}}}
\def\TriangleNumberTotalSNPos{N_{\text{\tiny $\ConnectivityMatrixSNPosVoid$}}}
\def\MaxSupportNbkpPosTriangles{{\cal T}_{N}^{+}(\sigma)}
\def\MaxSupportNbkpPosTrianglesVoid{{\cal T}_{N}^{+}}
\def\MaxSupportNbkpTriangles{{\cal T}_{N}(\sigma)}

\def\PointMatrixSN{{\cal P}_N(\sigma)}
\def\PointMatrixSNVoid{{\cal P}_N}
\def\ConnectivityMatrixSN{{\cal C}_N(\sigma)}
\def\ConnectivityMatrixSNVoid{{\cal C}_N}
\def\MaxSupportGbkpblTrangles{{\cal T}_{\gamma}(\sigma)}
\def\MaxSupportGbkpblTranglesVoid{{\cal T}_{\gamma}}

\def\ShapeFunction{\varphi}
\def\NodeSNPosNeg{N^{\pm}}
\def\TriangleSNPosNeg{T^{\pm}}


\def\CharacteristicFunctionTriangleSNPosi{\chi_{\TriangleSNPos_i}}
\def\CharacteristicFunctionTriangleSNPosj{\chi_{\TriangleSNPos_j}}
\def\CharacteristicFunctionTriangleSNPosk{\chi_{\TriangleSNPos_k}}
\def\CharacteristicFunctionTriangleSNPosl{\chi_{\TriangleSNPos_l}}
\def\CharacteristicFunctionTriangleSNPosm{\chi_{\TriangleSNPos_m}}

\def\CharacteristicFunctionTriangleSNNegi{\chi_{\TriangleSNNeg_i}}
\def\CharacteristicFunctionTriangleSNNegj{\chi_{\TriangleSNNeg_j}}
\def\CharacteristicFunctionTriangleSNNegk{\chi_{\TriangleSNNeg_k}}
\def\CharacteristicFunctionTriangleSNNegl{\chi_{\TriangleSNNeg_l}}
\def\CharacteristicFunctionTriangleSNNegm{\chi_{\TriangleSNNeg_m}}

\def\dW{\dot{\W}}
\def\dWImag{\dW^{(I)}}
\def\dWImagInit{\dW^{(I0)}}
\def\tWImag{\tW^{(I)}}
\def\tWImagInit{\dW^{(I0)}}

\def\tWImagAsy{\tW^{(Ia)}}

\def\CVgamma{\hat{\gamma}}
\def\CVgammaAsy{\CVgamma^{(a)}}

\def\deltak{\delta{k_2}}
\def\ThresholdK{K}
\def\Expal{E}
\def\DomainbkGreat{{\cal D}^c_0}

\def\EnergyDomain{{\cal D}_K(\sigma)}

\def\MPeakTwo{k_{2P}}
\def\MValleyTwo{k_{2V}}
\def\MPeakOne{k_{1\text{min}}}
\def\MValleyOne{k_{1\text{max}}}
\def\MMiddle{k_{20}(k_1;\sigma)}
\def\Supportgammabbbkp{{\cal S}_{\gamma^{(a)}_{22}}(\sigma)}
\def\Supportfbkp{{\cal S}_f(\sigma)}
\def\unitkone{\bar{k}_1}
\def\unitktwo{\bar{k}'_2}
\def\unitlone{\bar{l}_1}
\def\unitltwo{\bar{l}_2}

\def\MExtremeDomain{{\cal S}(\sigma)}
\def\MPeak{k_{2P}}
\def\MValley{k_{2V}}
\def\MMiddle{k_{20}(k_1;\sigma)}

\def\ba{\mathbf{a}}
\def\bb{\mathbf{b}}
\def\bc{\mathbf{c}}
\def\bd{\mathbf{d}}

\def\hr{\hat{r}}
\def\hs{\hat{s}}
\def\hbr{\hat{\br}}
\def\hbs{\hat{\bs}}

\def\ATriLinear{a^{\text{TL}}}
\def\ANodes{a^{(0)}}

\def\GOne{G_1}
\def\GFour{G_2}
\def\Gi{G_i}
\def\AAOne{D}
\def\Ai{D_i}
\def\AOne{D_1}
\def\ATwo{D_2}
\def\AThree{D_3}
\def\AFour{D_4}
\def\AFive{D_5}
\def\AOneAsy{\AOne^{(a)}}
\def\ATwoAsy{\ATwo^{(a)}}
\def\AThreeAsy{\AThree^{(a)}}
\def\AFourAsy{\AFour^{(a)}}
\def\AjAsy{D^{(a)}_j}
\def\ajAsy{d^{(a)}_j}
\def\aOneAsy{d^{(a)}_1}
\def\aTwoAsy{d^{(a)}_2}
\def\aThreeAsy{d^{(a)}_3}
\def\aFourAsy{d^{(a)}_4}

\def\GOneAsy{\GOne^{(a)}}
\def\GFourAsy{\GFour^{(a)}}
\def\AAsy{D^{(a)}}
\def\EAsy{\tW_{kk}^{(a)}}

\def\hk{\hat k}
\def\hl{\hat l}
\def\hbk{\hat\bk}
\def\hbl{\hat\bl}
\def\hm{\hat m}
\def\hbm{\hat\bm}

\def\maxk{\max\s{k}}

\def\GOne{G_1}
\def\GFour{G_2}
\def\AAOne{D}
\def\Ai{D_i}
\def\AOne{D_1}
\def\ATwo{D_2}
\def\AThree{D_3}
\def\AFour{D_4}
\def\AFive{D_5}
\def\AOneAsy{\AOne^{(a)}}
\def\ATwoAsy{\ATwo^{(a)}}
\def\AThreeAsy{\AThree^{(a)}}
\def\AFourAsy{\AFour^{(a)}}
\def\AjAsy{D^{(a)}_j}
\def\SupportAsy{{\cal D}^{(a)}}
\def\ajAsy{d^{(a)}_j}
\def\aOneAsy{d^{(a)}_1}
\def\aTwoAsy{d^{(a)}_2}
\def\aThreeAsy{d^{(a)}_3}
\def\aFourAsy{d^{(a)}_4}

\def\GOneAsy{\GOne^{(a)}}
\def\GFourAsy{\GFour^{(a)}}
\def\AAsy{D^{(a)}}
\def\EAsy{\tW_{kk}^{(a)}}

\def\hk{\hat k}
\def\hl{\hat l}
\def\hbk{\hat\bk}
\def\hbl{\hat\bl}
\def\hm{\hat m}
\def\hbm{\hat\bm}

 \def\hti{\hat t}
\def\hGOne{\hat{G}_1}
\def\hGFour{\hat G_2}
\def\hAAOne{\hat D}
\def\hAi{\hat D_i}
\def\hAOne{\hat D_1}
\def\hATwo{\hat D_2}
\def\hAThree{\hat D_3}
\def\hAFour{\hat D_4}
\def\hAFive{\hat D_5}

\def\GTriLinear{G^{\text{TL}}_2}
\def\GNodes{G^{(0)}_2}
\def\deltak{\delta \hk}
\def\tv{\omega}
\def\Vorticity{\omega}

\begin{abstract} 
We explore  incompressible homogeneous isotropic turbulence within the (fourth-order model)
formulation of optimal control and optimization, in contrast to the classical works of Proudman and 
Reid (1954) and Tatsumi (1957), with the intention to fix specially their defect of negative energy 
spectrum values being developed and to examine generally the conventional closure schemes. The 
isotropic forms for the general and spatially degenerated fourth order correlations of fluctuating 
velocity are obtained and the corresponding primary dynamical equations are derived. The degenerated 
fourth order correlation contains four scalar functions  $\Ai$, $i=1,2,3,4$, whose determination is 
the focus of closure. We discuss the constraints of equality for these functions as required by the 
self-consistency of the definition of the degenerated. Furthermore, we develop the constraints of 
inequality for the scalar functions based on the application of the Cauchy-Schwarz inequality, the 
non-negativity of the variance of 
products, and the non-negativity of the turbulent energy spectrum.
We intend to indicate the difficulty for a conventional scheme to satisfy all these constraints. As 
an alternative, we employ the turbulent energy per unit volume as the objective function to be 
maximized, under the constraints and the dynamical equations, with the four scalar functions as the 
control variables, which is a second-order cone programming problem. We then treat the asymptotic 
state solutions at large time and focus especially on the sub-model where the third order 
correlation is taken as the control variable, considering the computing resources available.
\end{abstract}

\section{Introduction}
\ \ \ \
In this part, we investigate  incompressible homogeneous isotropic turbulence
within the fourth-order model, 
in contrast to the classical works of Proudman and Reid \cite{ProudmanReid1954} 
and Tatsumi
\cite{Tatsumi1957};
 Our purpose is to resolve their flaw of negative energy spectrum values being developed
 (\cite{OBrienFrancis:1962}, \cite{Ogura:1963}).
We intend to demonstrate that,
though the simplest three-dimensional turbulent motion, 
  homogeneous isotropic turbulence provides us valuable information 
on why the  conventional turbulence modeling schemes have their defect  
and how statistical modeling should be framed.
Here, by the conventional, we mean that a closure is on the basis of certain equality
relations assumed 
among the correlations involved, especially between the highest order and the lower, 
like the quasi-normal adopted in \cite{ProudmanReid1954} and \cite{Tatsumi1957}.

Since we explore a different closure strategy,
 we need to develop the isotropic tensor representation for the general 
 fourth order correlation of fluctuating velocity
 $\overline{\wi(\bx) \wj(\by) \wk(\bz) \wl(\bz')}$,
under  the supposed incompressibility,  homogeneity, isotropy 
and the intrinsic symmetries from the correlation's definition.
Furthermore, we derive the isotropic tensor representation for
 the spatially degenerated  fourth order correlation
 $\overline{\wi(\bx) \wj(\bx) \wk(\by) \wl(\bz)}$,
considering that the evolution equation governing the third order correlation 
$\overline{\wi(\bx)\wj(\by) \wk(\bz)}$ involves only such a correlation 
and the computational complexity of the degenerated is much less than 
that of the general.
The interrelationship between the two is exploited
 to reduce the number of the scalar functions in the representation 
 for the degenerated,
 which are denoted as $\Ai$, $i=1,2,3,4$.
As a consequence of this representation, 
we need to re-derive the dynamical equations of evolution
for the two scalar functions contained in the well-known isotropic representation
 for $\overline{\wi(\bx) \wj(\by) \wk(\bz)}$ (\cite{ProudmanReid1954},
 \cite{Tatsumi1957}).

We discuss in detail the issues regarding the constraints.
To guarantee the self-consistency of the definition of the degenerated correlation,
 the above-mentioned representation needs to satisfy additional constraints
 under more degenerated conditions of spatial positions, such as
$$\overline{\wi(\bx) \wj(\bx) \wk(\by) \wl(\by)}
=\overline{\wk(\by) \wl(\by) \wi(\bx) \wj(\bx)}
$$
and
$$\overline{\wi(\bx) \wj(\bx) \wk(\bx) \wl(\by)}
\text{\ \ invariant under the permutation of $\{i$,\,$j$,\,$k\}$}
$$
These constraints of equality are expected to be satisfied through adequate structures
of $\Ai$.

There exist several sources of inequality constraints for the correlations. 
One results from the application of the Cauchy-Schwarz inequality to the correlations
and the structure functions
in the physical space. 
The second is from the requirement of
the non-negativity of the variance of products.
The third comes from the non-negativity of the turbulent energy spectrum.
These inequalities are, to a large extent, neglected by or even unenforcible
within the conventional schemes, except the limited implementation of realizability.
The satisfactions of these inequalities expectedly impose more
restrictions on the structures of $\Ai$,
 in addition to those from the constraints of equality.
 We should mention that the Cauchy-Schwarz inequality 
 and the non-negativity of the variance of products are natural parts of the
 closure strategy, because 
  the issue of closure in turbulence arises from the average treatment 
 of the Navier-Stokes equations
 and these inequalities are closely related to the mathematical ensemble average operation.
 
The above-mentioned  constraints are intrinsic to homogeneous isotropic turbulence,
 and their enforcement poses a great challenge to a conventional scheme, 
because such a scheme introduces a set
of equality relationships to represent the highest order correlations
 in terms of the correlations of lower orders, and these introduced or added
 may be incompatible with the intrinsic ones, as
demonstrated by the specific results of \cite{OBrienFrancis:1962} and \cite{Ogura:1963}.  
One strategy to accommodate all these intrinsic constraints is to adopt one
objective function to be optimized, constrained by the intrinsic constraints
 and the dynamical equations of evolution for the correlations,
 with  $\Ai$ as the control variables.
Therefore, the turbulence modeling problem is converted into an optimal control problem.
For homogeneous isotropic turbulence, we tentatively take the turbulent energy
per unit volume as the objective function to be maximized, and it is shown that
this mathematical formulation is a second-order cone programming (SOCP) problem when discretized.

We formulate the problem in both the physical and the Fourier wave-number spaces,
the latter makes it easy to represent the constraints and the objective function
explicitly in terms of the control variables, which offers us the advantage
to use the software packages available like `CVX/MOSEK' (\cite{CVX2011}, 
\cite{GrantBoyd2014}) to find numerical solutions.
The employment of Fourier transforms introduces higher-dimensional integrals, 
as a negative consequence.

We notice the challenge faced by the present optimal control strategy, 
such as the large size of numerical simulation resulting from the 
numerous constraints and finer discretization meshes,
which places much more demands on computing resources and algorithms.
Also, the issue of selection and redundancy of the constraints
and the issue of uniqueness of the solutions are not yet to be addressed.
More information will be gathered from the numerical simulation under consideration.

The present report is organized as follows.
We construct the mathematical structure for homogeneous isotropic turbulence in Section\,\ref{sec:BasicFormulation}.
As done conventionally, we start from the Navier-Stokes equations, introduce the correlations up to the fourth order
 and their symmetries, and present the equations of evolution for the correlations.
It is then followed by the Fourier transforms and the isotropic tensor representations of the correlations
 up to the fourth order in the Fourier wave-number space.
The primary dynamical equations for the scalar functions are derived, 
the constraints of equality and inequality are established either in the wave-number space or in the physical space, 
and the issue of objective functions is discussed.
Some general mathematical properties of the resultant SOCP problem are mentioned.
In Section\,\ref{sec:AsymptoticStateSolutions}, 
we treat the asymptotic state solutions at large time,
discussing certain possible characteristics and scaling.
Due to the restriction of computing resources, 
we focus on the sub-model where the third order correlation is taken
as the control variable.

\section{Basic Formulation}\label{sec:BasicFormulation}
\ \ \ \
Let us consider  homogeneous isotropic turbulence in $\mathbb{R}^3$.
 The fluctuation fields of velocity $\wi(\bx, t)$ and (scaled) pressure $\q(\bx,t):=p(\bx,t)/\rho$
 are supposedly governed by the incompressible Navier-Stokes equations, 
\begin{align}
\frac{\partial\wk}{\partial x_k}=0,\quad
\frac{\partial \wi}{\partial t}
+\frac{\partial (\wi \wk)}{\partial x_k}
=
-\frac{\partial \q}{\partial x_i}
+\nu \frac{\partial^2 \wi}{\partial x_k \partial x_k},\quad
\frac{\partial^2 \q}{\partial x_k \partial x_k}
=
-\frac{\partial^2 (\wl \wk)}{\partial x_k \partial x_l}
\label{HIT_NSEqsInPhysicalSpace}
\end{align}
we can then construct the following equations for the evolution of the multi-point (tensor) correlations up to the fourth order,
\begin{align}
&
\frac{\partial}{\partial x_k}\overline{\wk(\bx) \wj(\by)}=0,\quad
\frac{\partial}{\partial x_k}\overline{\wk(\bx) \wj(\by) \wl(\bz)}=0,\quad
\frac{\partial}{\partial y_k}\overline{\wi(\bx) \wj(\bx) \wk(\by) \wl(\bz)}=0,
\notag\\[4pt]
&
\frac{\partial}{\partial x_i}\overline{\wi(\bx) \wj(\by) \wk(\bz) \wl(\bz')}=0,
\quad
\frac{\partial}{\partial x_k}\overline{\wk(\bx) \q(\by)}=0,\quad
\frac{\partial}{\partial x_k}\overline{\wk(\bx) \wl(\by) \q(\bz)}=0
\label{HIT_DivergenceFreeInPhysicalSpace_2p_3p}
\end{align}
\begin{align}
&
\frac{\partial}{\partial t}\overline{\wi(\bx) \wj(\by)}
+\frac{\partial}{\partial x_k}\overline{\wi(\bx) \wk(\bx) \wj(\by)}
+\frac{\partial}{\partial y_k}\overline{\wj(\by) \wk(\by) \wi(\bx)}
\notag\\[4pt]
=&
-\frac{\partial }{\partial x_i}\overline{\q(\bx) \wj(\by)}
-\frac{\partial }{\partial y_j}\overline{\q(\by) \wi(\bx)}
+ \nu\,\bigg(\frac{\partial^2}{\partial x_k \partial x_k}+\frac{\partial^2}{\partial y_k\partial y_k}\bigg)
\overline{\wi(\bx) \wj(\by)}
\label{HIT_CLMInPhysicalSpace_2p}
\end{align}
\begin{align}
&
 \frac{\partial }{\partial t}\overline{\wi(\bx)\wj(\by)\wk(\bz)}
+\frac{\partial }{\partial x_l}\overline{\wi(\bx) \wl(\bx)\wj(\by)\wk(\bz)}
\notag\\[4pt]&
+\frac{\partial }{\partial y_l}\overline{\wj(\by) \wl(\by)\wi(\bx)\wk(\bz)}
+\frac{\partial }{\partial z_l}\overline{\wk(\bz) \wl(\bz)\wi(\bx)\wj(\by)}
\notag\\[4pt]
=\,&
-\frac{\partial }{\partial x_i}\overline{\q(\bx) \wj(\by)\wk(\bz)}
-\frac{\partial }{\partial y_j}\overline{\q(\by) \wi(\bx)\wk(\bz)}
-\frac{\partial }{\partial z_k}\overline{\q(\bz) \wi(\bx)\wj(\by)}
\notag\\[4pt]&
+\nu\, \bigg(
 \frac{\partial^2 }{\partial x_l\partial x_l}
+\frac{\partial^2 }{\partial y_l\partial y_l}
+\frac{\partial^2 }{\partial z_l\partial z_l}\bigg)\overline{\wi(\bx)\wj(\by)\wk(\bz)}
\label{HIT_CLMInPhysicalSpace_3p}
\end{align}
\begin{align}
\frac{\partial^2 }{\partial x_k x_k}\overline{\q(\bx) \wj(\by)}
=
-\frac{\partial^2}{\partial x_k \partial x_l}\overline{\wl(\bx) \wk(\bx) \wj(\by)}
\label{HIT_PressureInPhysicalSpace_2p}
\end{align}
\begin{align}
\frac{\partial^2 }{\partial x_l x_l}\overline{\q(\bx)\wj(\by)\wk(\bz)}
=
-\frac{\partial^2 }{\partial x_m \partial x_l}\overline{\wl(\bx) \wm(\bx)\wj(\by)\wk(\bz)}
\label{HIT_PressureInPhysicalSpace_3p}
\end{align}
and
\begin{align}
\frac{\partial^2 }{\partial y_k\partial y_k}\overline{\q(\bx)\,\q(\by)}
=
-\frac{\partial^2}{\partial y_k \partial y_l}\overline{\q(\bx) \wk(\by) \wl(\by)}
\label{HIT_PressureInPhysicalSpace_qq_2p}
\end{align}

Following the conventional treatment of homogeneity (\cite{ProudmanReid1954}, \cite{Tatsumi1957}), we adopt
\begin{align}
&
\Wij(\br):=\overline{\wi(\bx) \wj(\by)}=\overline{\wi(\mathbf{0}) \wj(\br)},\quad 
\W_{ijk}(\br,\bs):=\overline{\wi(\bx) \wj(\by) \wk(\bz)}=\overline{\wi(\mathbf{0}) \wj(\br) \wk(\bs)},
\notag\\[4pt]
&
\W_{(ij)kl}(\br,\bs):=\overline{\wi(\bx) \wj(\bx) \wk(\by) \wl(\bz)}
=\overline{\wi(\mathbf{0}) \wj(\mathbf{0}) \wk(\br) \wl(\bs)},
\notag\\[4pt]
&
\W_{ijkl}(\br,\bs,\bs'):=\overline{\wi(\bx) \wj(\by) \wk(\bz) \wl(\bz')}=\overline{\wi(\mathbf{0}) \wj(\mathbf{\br}) \wk(\bs) \wl(\bs')},
\notag\\[4pt]
&
\Q(\br):=\overline{\q(\bx) \,\q(\by)}=\overline{\q(\mathbf{0})\, \q(\br)},
\quad
\Qj(\br):=\overline{\q(\bx) \wj(\by)}=\overline{\q(\mathbf{0}) \wj(\br)},
\notag\\[4pt]
&
\Q_{jk}(\br,\bs):=\overline{\q(\bx)\wj(\by)\wk(\bz)}=\overline{\q(\mathbf{0})\wj(\br)\wk(\bs)}
\label{Homogeneity}
\end{align}
Here, $\br:=\by-\bx$, $\bs:=\bz-\bx$ and $\bs':=\bz'-\bx$.
The dependence of the correlations on $t$ is suppressed for the sake of brevity.
For the fourth order correlation of velocity fluctuations,
 we include both the general $\W_{ijkl}(\br,\bs,\bs')$ and the spatially degenerated $\W_{(ij)kl}(\br,\bs)$,
 whose consequences are to be explored later.
The definitions of \eqref{Homogeneity} result in the symmetry properties of
\begin{align}
&
\Wij(\br)=\Wji(-\br),\quad
\W_{ijk}(\br,\bs)=\W_{ikj}(\bs,\br)=\W_{jik}(-\br,\bs-\br)=\W_{kij}(-\bs,\br-\bs),
\notag\\[4pt]&
 \W_{(ij)kl}(\br,\bs)=\W_{(ji)kl}(\br,\bs)=\W_{(ij)lk}(\bs,\br),\quad
 \W_{(ij)kl}(\br,\br)=\W_{(kl)ij}(-\br,-\br),
\notag\\[4pt]&
\W_{(ij)kl}(\bo,\br)=\W_{(ik)jl}(\bo,\br),\quad
\W_{(ij)kl}(\bo,\bo)\ \ \text{invariant under permutation of $\{i,j,k,l\}$},
\notag\\[4pt]
&
\W_{ijkl}(\br,\bs,\bs')=\W_{ijlk}(\br,\bs',\bs)=\W_{ilkj}(\bs',\bs,\br)=\W_{ikjl}(\bs,\br,\bs')
=\W_{jikl}(-\br,\bs-\br,\bs'-\br)
\notag\\[4pt]
&
=\W_{kijl}(-\bs,\br-\bs,\bs'-\bs)=\W_{lijk}(-\bs',\br-\bs',\bs-\bs'),
\quad
\W_{(ij)kl}(\br,\bs)=\W_{ijkl}(\bo,\br,\bs),
\notag\\[4pt]
&
\Q(\br)=\Q(-\br),
\quad
\Q_{jk}(\br,\bs)=\Q_{kj}(\bs,\br)
\label{Homogeneity_Symmetry}
\end{align}

Considering the zero average velocity field, we also impose the inversion symmetry of
\begin{align}
&
\W_{ij}(\br)=\W_{ij}(-\br),\quad
\W_{ijk}(\br,\bs)=-\W_{ijk}(-\br,-\bs),\quad
\W_{(ij)kl}(\br,\bs)=\W_{(ij)kl}(-\br,-\bs),
\notag\\[4pt]&
\W_{ijkl}(\br,\bs,\bs')=\W_{ijkl}(-\br,-\bs,-\bs'),\quad
\Q(\br)=\Q(-\br),\quad
\Q_j(\br)=-\Q_j(-\br),
\notag\\[4pt]&
\Q_{jk}(\br,\bs)=\Q_{jk}(-\br,-\bs)
\label{Homogeneity_Inversion}
\end{align}

With the help of \eqref{Homogeneity_Symmetry} and \eqref{Homogeneity_Inversion},
we insert \eqref{Homogeneity} into 
\eqref{HIT_DivergenceFreeInPhysicalSpace_2p_3p} through \eqref{HIT_PressureInPhysicalSpace_qq_2p} to get
\begin{align}
&
\frac{\partial}{\partial r_k}\W_{kj}(\br)=0,\ \
\bigg(\frac{\partial}{\partial r_k}+\frac{\partial}{\partial s_k}\bigg)\W_{kjl}(\br,\bs)=0,\ \
\frac{\partial}{\partial r_j}\W_{kjl}(\br,\bs)=0,\ \
\frac{\partial}{\partial r_k}\W_{(ij)kl}(\br,\bs)=0,
\notag\\[4pt]&
\bigg(\frac{\partial}{\partial r_i}+\frac{\partial}{\partial s_i}+\frac{\partial}{\partial s'_i}\bigg)
\W_{ijkl}(\br,\bs,\bs')=0,\ \
\frac{\partial}{\partial r_j}\W_{ijkl}(\br,\bs,\bs')=0,\ \
\frac{\partial}{\partial r_k}\Q_{k}(\br)=0,
\notag\\[4pt] &
\frac{\partial}{\partial r_k}\Q_{kl}(\br,\bs)=0
\label{HIT_DivergenceFreeInPhysicalSpace_2p_3p_rs}
\end{align}
\begin{align}
\frac{\partial}{\partial t}\W_{ij}(\br)
-\frac{\partial}{\partial r_k}\W_{ikj}(\mathbf{0},\br)
+\frac{\partial}{\partial r_k}\W_{jki}(\mathbf{0},-\br)
= \frac{\partial }{\partial r_i}\Q_j(\br)
-\frac{\partial }{\partial r_j}\Q_i(-\br)
+2 \,\nu\, \frac{\partial^2}{\partial r_k\partial  r_k}\W_{ij}(\br)
\label{HIT_CLMInPhysicalSpace_2p_r}
\end{align}
\begin{align}
&
 \frac{\partial }{\partial t} \W_{ijk}(\br,\bs)            
-\bigg(\frac{\partial }{\partial r_l}+\frac{\partial }{\partial s_l}\bigg) \W_{(il)jk}(\br,\bs) 
+\frac{\partial }{\partial r_l} \W_{(jl)ik}(-\br,\bs-\br)    
+\frac{\partial }{\partial s_l} \W_{(kl)ij}(-\bs,\br-\bs)       
\notag\\[4pt]
=\,&
 \frac{\partial }{\partial r_i} \Q_{jk}(\br,\bs)  
+\frac{\partial }{\partial s_i} \Q_{jk}(\br,\bs)  
-\frac{\partial }{\partial r_j} \Q_{ik}(-\br,\bs-\br)   
-\frac{\partial }{\partial s_k} \Q_{ij}(-\bs,\br-\bs)         
\notag\\[4pt]&
+2\,\nu\, \bigg(
 \frac{\partial^2 }{\partial r_l\partial  r_l}
+\frac{\partial^2 }{\partial s_l\partial  s_l}
+\frac{\partial^2}{\partial r_l \partial s_l}
\bigg) \W_{ijk}(\br,\bs)  
\label{HIT_CLMInPhysicalSpace_3p_rs}
\end{align}
\begin{align}
\frac{\partial^2 }{\partial r_k\partial  r_k} \Qj(\br) 
= 
-\frac{\partial^2}{\partial r_k \partial r_l} \W_{lkj}(\mathbf{0},\br)
\label{HIT_PressureInPhysicalSpace_2p_r}
\end{align}
\begin{align}
\bigg(\frac{\partial}{\partial r_l}+\frac{\partial}{\partial s_l}\bigg)\bigg(\frac{\partial}{\partial r_l}
+\frac{\partial}{\partial s_l}\bigg)\Q_{jk}(\br,\bs) 
=
-\bigg(\frac{\partial}{\partial r_m}+\frac{\partial}{\partial s_m}\bigg)\bigg(\frac{\partial}{\partial r_l}
+\frac{\partial}{\partial s_l}\bigg) \W_{(lm)jk}(\br,\bs)
\label{HIT_PressureInPhysicalSpace_3p_rs}
\end{align}
and
\begin{align}
\frac{\partial^2 }{\partial r_k r_k}\Q(\br)
=
-\frac{\partial^2}{\partial r_k \partial r_l}\Q_{kl}(\br,\br)
\label{HIT_PressureInPhysicalSpace_qq_2p_r}
\end{align}

\subsection{Fourier Transforms}
\ \ \ \
It is convenient to reformulate the above mathematical relations with the help of Fourier transforms.
The treatment converts the partial spatial derivatives into algebraic operations
 and introduces the turbulent energy spectrum in the Fourier wave-number space.
The treatment makes it easy to formulate the constraints and the objective function explicitly 
in terms of the control variables which helps to solve the problem numerically, as to be discussed.
Its negative side is that higher-dimensional integrals are involved in the formulation as indicated
 by \eqref{FourierTransform} below.

 We employ the Fourier transforms,
\begin{align}
&
\Wij(\br)=\int_{\mathbb{R}^3} d\bk\, \tWij(\bk)\, \exp(\imaginary\, \bk\s\cdot\s\br),
\quad
\W_{ijk}(\br,\bs)=\int_{\mathbb{R}^3\times\mathbb{R}^3} d\bk\, d\bl\,
          \tW_{ijk}(\bk,\bl)\, \exp\s\l[\imaginary\, (\bk\s\cdot\s\br+\bl\s\cdot\s\bs)\r],
\notag\\[4pt]&
\W_{(ij)kl}(\br,\bs)=
\int_{\mathbb{R}^3\times\mathbb{R}^3} d\bk\, d\bl\,
 \tW_{(ij)kl}(\bk,\bl)\, \exp\s\l[\imaginary\, (\bk\s\cdot\s\br+\bl\s\cdot\s\bs)\r],
\notag\\[4pt]&
\W_{ijkl}(\br,\bs,\bs')=\int_{\mathbb{R}^3\times\mathbb{R}^3\times\mathbb{R}^3} d\bk\, d\bl\, d\bm\,
 \tW_{ijkl}(\bk,\bl,\bm)\, \exp\s\l[\imaginary\, (\bk\s\cdot\s\br+\bl\s\cdot\s\bs+\bm\s\cdot\s\bs')\r],
\notag\\[4pt]&
\Q(\br)=\int_{\mathbb{R}^3} d\bk\, \tQ(\bk)\, \exp(\imaginary\, \bk\s\cdot\s\br),
\quad 
\Qj(\br)=\int_{\mathbb{R}^3} d\bk\, \tQj(\bk)\, \exp(\imaginary\, \bk\s\cdot\s\br),
\notag\\[4pt]&
\Q_{jk}(\br,\bs)=\int_{\mathbb{R}^3\times\mathbb{R}^3} d\bk \,d\bl\,
       \tQ_{jk}(\bk,\bl)\,\exp\s\l[\imaginary\, (\bk\s\cdot\s\br+\bl\s\cdot\s\bs)\r]
\label{FourierTransform}
\end{align}
That the correlations in the physical space are real requires that
\begin{align}
&
\tW^*_{ij}(\bk) =\tW_{ij}(-\bk),
\quad
\tW^*_{ijk}(\bk,\bl) =\tW_{ijk}(-\bk,-\bl),
\quad
\tW^*_{(ij)kl}(\bk,\bl) =\tW_{(ij)kl}(-\bk,-\bl),
\notag\\[4pt]&
\tW^*_{ijkl}(\bk,\bl,\bm) =\tW_{ijkl}(-\bk,-\bl,-\bm),\quad
\tQ^*(\bk)=\tQ(-\bk),
\quad
\tQ^*_j(\bk)=\tQ_j(-\bk),
\notag\\[4pt]&
\tQ^*_{jk}(\bk,\bl)=\tQ_{jk}(-\bk,-\bl)
\label{RealCorrelations_Inps}
\end{align}
where the superscript $*$ denotes the complex conjugate operation.

Substituting \eqref{FourierTransform} into \eqref{Homogeneity_Symmetry} and 
\eqref{Homogeneity_Inversion} and
combining with \eqref{RealCorrelations_Inps}, we obtain
\begin{align}
&
\tW_{ij}(\bk)=\tW_{ji}(\bk)=\tW_{ij}(-\bk)=\tW^*_{ij}(\bk),
\notag\\[4pt]
&
\tW_{ijk}(\bk,\bl)=\tW_{ikj}(\bl,\bk)=\tW_{jik}(-\bk-\bl,\bl)
=\tW_{kij}(-\bk-\bl,\bk)=-\tW_{ijk}(-\bk,-\bl)=-\tW^*_{ijk}(\bk,\bl),
\notag\\[4pt]
&
 \tW_{(ij)kl}(\bk,\bl)=\tW_{(ji)kl}(\bk,\bl)=\tW_{(ij)lk}(\bl,\bk)
=\tW_{(ij)kl}(-\bk,-\bl)
=\tW^*_{(ij)kl}(\bk,\bl),
\notag\\[4pt]&
\int_{\mathbb{R}^3} d\bl
 \l[\tW_{(ij)kl}(\bk+\bl,-\bl)-\tW_{(kl)ij}(-\bk-\bl,\bl)\r]=0,
\quad
\int_{\mathbb{R}^3} d\bl
 \l[\tW_{(ij)kl}(\bl,\bk)-\tW_{(ik)jl}(\bl,\bk)\r]=0,
\notag\\[4pt]
&
\tW_{ijkl}(\bk,\bl,\bm)=\tW_{ijlk}(\bk,\bm,\bl)
=\tW_{ilkj}(\bm,\bl,\bk)=\tW_{ikjl}(\bl,\bk,\bm)
=\tW_{jikl}(-\bk-\bl-\bm,\bl,\bm)
\notag\\[4pt]
&
=\tW_{kijl}(-\bk-\bl-\bm,\bk,\bm)
=\tW_{lijk}(-\bk-\bl-\bm,\bk,\bl)
=\tW_{ijkl}(-\bk,-\bl,-\bm)=\tW^*_{ijkl}(\bk,\bl,\bm),
\notag\\[4pt]
&
\tQ(\bk)=\tQ(-\bk)=\tQ^*(\bk),
\quad
\tQ_j(\bk)=-\tQ_j(-\bk)=-\tQ^*_j(\bk),
\notag\\[4pt]
&
\tQ_{jk}(\bk,\bl)=\tQ_{kj}(\bl,\bk)=\tQ_{jk}(-\bk,-\bl)=\tQ^*_{jk}(\bk,\bl)
\label{Homogeneity_Symmetry_Inversion_fs}
\end{align}
which indicate that $\tW_{ij}(\bk)$, $\tW_{(ij)kl}(\bk,\bl)$,
 $\tW_{ijkl}(\bk,\bl,\bm)$, $\tQ(\bk)$ and $\tQ_{ij}(\bk,\bl)$ are real
 and $\tW_{ijk}(\bk,\bl)$ and $\tQ_j(\bk)$ are purely imaginary.
Next,
substitution of \eqref{FourierTransform} into 
\eqref{HIT_DivergenceFreeInPhysicalSpace_2p_3p_rs} through
 \eqref{HIT_PressureInPhysicalSpace_qq_2p_r} results in
\begin{align}
&
k_k\,\tW_{kj}(\bk) =0,\ \
 \big(k_k+l_k\big)\,\tW_{kjl}(\bk,\bl)=0,\ \
k_j\,\tW_{kjl}(\bk,\bl)=0,\ \
 k_k\,\tW_{(ij)kl}(\bk,\bl)=0,
\notag\\[4pt] &
(k_i+l_i+m_i)\,\tW_{ijkl}(\bk,\bl,\bm)=0,\ \
k_j\,\tW_{ijkl}(\bk,\bl,\bm)=0,\ \
k_k\,\tQ_{k}(\bk)=0,\ \
k_k\,\tQ_{kl}(\bk,\bl)=0
\label{HIT_DivergenceFreeInPhysicalSpace_qandw_fs}
\end{align}
\begin{align}
 \tQ(\bk) 
=
-\frac{k_k\,k_l}{|\bk|^2}\,\int_{\mathbb{R}^3} d\bl\,\tQ_{kl}(\bk-\bl,\bl)
\label{HIT_PressureInPhysicalSpace_qq_fs}
\end{align}
\begin{align}
\tQ_j(\bk) = 
-\frac{k_l\,k_k}{|\bk|^2}\,\int_{\mathbb{R}^3} d\bl \, \tW_{lkj}(\bl,\bk)
\label{HIT_PressureInPhysicalSpace_qw_fs}
\end{align}
\begin{align}
\tQ_{jk}(\bk,\bl)\,
=
-\frac{(k_l+l_l)\,(k_m+l_m)}{|\bk+\bl|^2}\,\tW_{(lm)jk}(\bk,\bl)           
\label{HIT_PressureInPhysicalSpace_qww_fs}
\end{align}
\begin{align}
 \frac{\partial}{\partial t}\tW_{ij}(\bk)    
+2\,\nu\,|\bk|^2\,\tW_{ij}(\bk)   
=
 \imaginary\, k_i\,\tQj(\bk)- \imaginary\, k_j\,\tQi(-\bk) 
+\imaginary\,k_k \int_{\mathbb{R}^3} d\bl\,
\Big(\tW_{ijk}(\bk,\bl)-\tW_{jik}(-\bk,\bl)\Big)  
\label{HIT_CLMInPhysicalSpace_ww_fs}
\end{align}
and
\begin{align}
&
 \frac{\partial }{\partial t}\tW_{ijk}(\bk,\bl)           
+\nu\,\big(|\bk|^2+|\bl|^2+|\bk+\bl|^2\big)\,\tW_{ijk}(\bk,\bl) 
\notag\\[4pt]
=&\,
 \imaginary\,(k_i+l_i)\,\tQ_{jk}(\bk,\bl)   
-\imaginary\,k_j\,\tQ_{ik}(-\bk-\bl,\bl) 
-\imaginary\,l_k\,\tQ_{ij}(-\bk-\bl,\bk)       
\notag\\[4pt] &     
+\imaginary\, (k_l+l_l) \,\tW_{(il)jk}(\bk,\bl) 
-\imaginary\,k_l\,\tW_{(jl)ik}(-\bk-\bl,\bl)
-\imaginary\,l_l \,\tW_{(kl)ij}(-\bk-\bl,\bk)     
\label{HIT_CLMInPhysicalSpace_www_fs}
\end{align}

In the Fourier wave-number space,
 the general and the degenerated fourth order correlations are related through
\begin{align}
 \tW_{(ij)kl}(\bk,\bl)=\int_{\mathbb{R}^3} d\bm\, \tW_{ijkl}(\bm,\bk,\bl)
\label{tW(ijkl)VstWijkl}
\end{align}
following from the link in \eqref{Homogeneity_Symmetry},
whose consequences will be explored in Subsection\,\ref{Subsec:RelationshipBetween}.
Equations \eqref{HIT_CLMInPhysicalSpace_www_fs}, \eqref{HIT_PressureInPhysicalSpace_qww_fs}
and \eqref{tW(ijkl)VstWijkl}
indicate that the degenerated $\tW_{(IJ)KL}$ plays the role of an 
intermediate variable between the rate of change of $\tW_{ijk}$
and the general $\tW_{IJKL}$, if the latter is taken as the control variable.

\subsection{Isotropy}
\ \ \ \
According to \cite{ProudmanReid1954},
 it is sufficient to formulate the isotropic forms of the tensor correlations
 in the Fourier wave-number space 
under the supposed isotropy, along with the constraints of
\eqref{Homogeneity_Symmetry_Inversion_fs} and 
\eqref{HIT_DivergenceFreeInPhysicalSpace_qandw_fs}.
We have the well-known
\begin{align}
\tQ_j(\bk)=0
\label{tQj_Isotropic}
\end{align}
\vskip-4mm
\begin{align}
\tW_{ij}(\bk)
=
 \frac{1}{2}\,\bigg(\delta_{ij}-\frac{k_i\,k_j}{k^2}\bigg)\,\tW_{kk}(k)
\label{tWij_Isotropic}
\end{align}
and from \cite{ProudmanReid1954},
\begin{align}
 \tW_{ijk}(\bk,\bl)=\,&
\imaginary\,\Delta_{im}(\bm)\,\Delta_{jn}(\bk)\,\Delta_{kp}(\bl)
\notag\\[4pt]&\hskip5mm\times
\Big(
 k_m\, l_n\, k_p\, \GOne(m,k,l)
+\delta_{mn}\,k_p\,\GFour(m,l,k)
+\delta_{np}\,k_m\,\GFour(l,m,k)
\notag\\[4pt]&\hskip12mm
+\delta_{pm}\,l_n\,\GFour(m,k,l)
\Big),\quad \bk+\bl+\bm=\bo
\label{tWijk_Isotropic}
\end{align}
where $\Delta_{im}(\bm):=\delta_{im}-m_i\,m_m/m^2$ is the second order tensor 
to enforce the divergence-free condition from the incompressibility, 
the two scalar functions contained in \eqref{tWijk_Isotropic} are real 
and possess the symmetry properties of 
\begin{align}
\GOne(m,k,l)=-\GOne(m,l,k)=-\GOne(k,m,l),\ \
\GFour(m,k,l)=-\GFour(l,k,m),\ \ \bk+\bl+\bm=\bo
\label{tWijk_Isotropic_Constraints}
\end{align}

For the fourth order correlation within the fourth-order model,
 we encounter two possible representations, the general $\tW_{ijkl}(\bk,\bl,\bm)$ and the degenerated $\tW_{(ij)kl}(\bk,\bl)$,
 and we need to derive their isotropic forms, considering that we explore a closure scheme
 different from the conventional, such as the quasi-normal or its variants.
Obviously, equations \eqref{HIT_PressureInPhysicalSpace_qq_fs} through \eqref{HIT_CLMInPhysicalSpace_www_fs}
 involve the degenerated but not the general, which lends one basis to employ only $\tW_{(ij)kl}(\bk,\bl)$.
However, the isotropic form of $\tW_{ijkl}(\bk,\bl,\bm)$ provides certain restrictions
 on the structure of $\tW_{(ij)kl}(\bk,\bl)$ and vice versa,
 and it also offers certain clarities on the issue of closure, 
as to be shown.

We outline the major steps to determine the isotropic form of $\tW_{ijkl}(\bk,\bl,\bm)$ below.

We find first a primary general fourth-order tensor function $\Phi_{ijkl}(\bk,\bl,\bm)$
 of wave-number vectors $\bk$, $\bl$ and $\bm$,
which contains the arbitrary scalar functions of the invariants of the wave-numbers,
 $k$, $l$, $m$, $|\bk+\bl|$, $|\bl+\bm|$ and $|\bm+\bk|$.
Next, we need to impose the symmetry properties of $\tW_{ijkl}(\bk,\bl,\bm)$
 listed in \eqref{Homogeneity_Symmetry_Inversion_fs} to restrict the structure of $\Phi_{ijkl}$;
 the related algebraic operations are quite complicated 
 since $\Phi_{ijkl}$ contains a very large number of terms.
To avoid this complicity, we take a different route.
Specifically, considering that $\tW_{ijkl}$ needs to satisfy
 the divergence-free conditions of \eqref{HIT_DivergenceFreeInPhysicalSpace_qandw_fs}
we employ the projection \cite{ProudmanReid1954}
\begin{align}
\tW_{IJKL}(\bk,\bl,\bm)
=\Delta_{Ii}(\bn)\,\Delta_{Jj}(\bk)\,\Delta_{Kk}(\bl)\,\Delta_{Ll}(\bm)\,\Phi_{ijkl}(\bk,\bl,\bm),
\quad
\bn=\bk+\bl+\bm
\end{align}
The multiplication by $\Delta_{Ii}(\bn)\cdots\Delta_{Ll}(\bm)$ eliminates many of the terms 
and the scalar functions contained in $\Phi_{ijkl}$ to reduce the above expression to
\begin{align}
\tW_{IJKL}(\bk,\bl,\bm)
=\Delta_{Ii}(\bn)\,\Delta_{Jj}(\bk)\,\Delta_{Kk}(\bl)\,\Delta_{Ll}(\bm)\,\Phi'_{ijkl}(\bk,\bl,\bm),
\quad
\bn=\bk+\bl+\bm
\label{WIJKL_DivergenceFreeImplementation}
\end{align}
We then impose the symmetry properties of $\tW_{ijkl}$ directly on the reduced $\Phi'_{ijkl}$. 
The resultant $\tW_{IJKL}$ of \eqref{WIJKL_DivergenceFreeImplementation}
satisfies these symmetry properties too, since it can be verified that 
\eqref{WIJKL_DivergenceFreeImplementation} and \eqref{Homogeneity_Symmetry_Inversion_fs} lead to
\begin{align}
&
 \Delta_{Ii}(\bn)\cdots\Delta_{Ll}(\bm)\,\Phi'_{ijkl}(\bk,\bl,\bm)
=\Delta_{Ii}(\bn)\cdots\Delta_{Ll}(\bm)\,\Phi'_{ijlk}(\bk,\bm,\bl)
\notag\\[4pt]&
=\Delta_{Ii}(\bn)\cdots\Delta_{Ll}(\bm)\,\Phi'_{ilkj}(\bm,\bl,\bk)
=\Delta_{Ii}(\bn)\cdots\Delta_{Ll}(\bm)\,\Phi'_{ikjl}(\bl,\bk,\bm)
\notag\\[4pt]&
=\Delta_{Ii}(\bn)\cdots\Delta_{Ll}(\bm)\,\Phi'_{jikl}(-\bk-\bl-\bm,\bl,\bm)
=\Delta_{Ii}(\bn)\cdots\Delta_{Ll}(\bm)\,\Phi'_{kijl}(-\bk-\bl-\bm,\bk,\bm)
\notag\\[4pt]&
=\Delta_{Ii}(\bn)\cdots\Delta_{Ll}(\bm)\,\Phi'_{lijk}(-\bk-\bl-\bm,\bk,\bl)
=\Delta_{Ii}(\bn)\cdots\Delta_{Ll}(\bm)\,\Phi'_{ijkl}(-\bk,-\bl,-\bm)
\end{align}
Each implementation of the symmetry properties on $\Phi'_{ijkl}$ results in
 a fourth-order tensor equation, a summation
 of the elements like $k_i l_j k_k k_l$ 
 with their coefficients composed of the scalar functions in $\Phi'_{ijkl}$.
That is, the equation is a linear algebraic equality of  multivariate polynomials of wave-number vector components.
For the sake of simplicity,
 we take the standard form of multivariate polynomials as the basis and use its linear independence to infer 
the values of the coefficients involved.
To make the derivation procedure simpler, 
we start with those symmetries in \eqref{Homogeneity_Symmetry_Inversion_fs}
 containing the argument $-\bk-\bl-\bm$
 and update the form of $\Phi'_{ijkl}$ sequentially.

We may also follow the procedure suggested in \cite{ProudmanReid1954},
by applying the symmetries directly to  $\tW_{IJKL}$ of \eqref{WIJKL_DivergenceFreeImplementation} 
so as to determine $\Phi'_{ijkl}$.
Here, we face the challenge to deal with the large number of terms due to the expansion of 
$\Delta_{Ii}(\bn)\cdots\Delta_{Ll}(\bm)$.
Whether the two approaches produce the same result is yet to be resolved.

With a lengthy procedure, we obtain finally
\begin{align}
\tW_{IJKL}(\bk,\bl,\bm)
=\,&
\Delta_{Ii}(\bn)\,
\Delta_{Jj}(\bk)\,
\Delta_{Kk}(\bl)\,
\Delta_{Ll}(\bm)\,
\notag\\[4pt]&
\times
\Big(
 \delta_{ij}\,\delta_{kl}\,\AAOne(\bk,\bl,\bm)
+\delta_{ik}\,\delta_{jl}\,\AAOne(\bl,\bk,\bm)
+\delta_{il}\,\delta_{jk}\,\AAOne(\bm,\bl,\bk)
\Big),\ \bn=\bk+\bl+\bm
\label{tWijkl_Isotropic}
\end{align}
Here, the scalar function $\AAOne(\bk,\bl,\bm)$ possesses the symmetry properties of
\begin{align}
 \AAOne(\bk,\bl,\bm)
=\AAOne(\bk,\bm,\bl)
=\AAOne(\bm,\bk,-\bk-\bl-\bm)
=\AAOne(\bl,\bk,-\bk-\bl-\bm)
\label{tWijkl_Isotropic_Coefficient}
\end{align}
with $(\bk,\bl,\bm)$ denoting the relevant invariants, $(k$, $l$, $m$, $|\bk+\bl|$, $|\bl+\bm|$,
 $|\bm+\bk|)$. These invariants will be restricted further to $k$, $l$, $m$ and $|\bk+\bl+\bm|$
 by \eqref{tW(ijkl)VstWijkl},
as to be demonstrated in Subsection \ref{Subsec:RelationshipBetween}.

In the degenerated case of $\tW_{(ij)kl}(\bk,\bl)$,
 we follow the idea similar to the above, enforcing the associated constraints 
 of the non-integral form in \eqref{Homogeneity_Symmetry_Inversion_fs} and in \eqref{HIT_DivergenceFreeInPhysicalSpace_qandw_fs}
 to infer that
\begin{align}
&
\tW_{(IJ)KL}(\bk,\bl)
\notag\\
=\,&
\Delta_{Kk}(\bk)\,
\Delta_{Ll}(\bl)\,
\Big[
 \delta_{IJ}\,\delta_{kl}\,\AOne(k,l,|\bk+\bl|)
+(\delta_{Ik}\,\delta_{Jl}+\delta_{Il}\,\delta_{Jk})\,\ATwo(k,l,|\bk+\bl|)
\notag\\[4pt]&\hskip27mm
+\delta_{IJ}\,l_k\,k_l\,\AThree(k,l,|\bk+\bl|)
+(l_I\,k_J+k_I\,l_J)\,l_k\,k_l\,\AFour(k,l,|\bk+\bl|)
\notag\\[4pt]&\hskip27mm
+k_I\,k_J\,l_k\,k_l\,\AFive(k,l,|\bk+\bl|)
+l_I\,l_J\,l_k\,k_l\,\AFive(l,k,|\bk+\bl|)
\Big]
\label{tW(ij)kl_IsotropicA}
\end{align}
where the scalar functions have the symmetry properties of
\begin{align}
\Ai(k,l,|\bk+\bl|)=\Ai(l,k,|\bl+\bk|),\ \ i=1,2,3,4
\label{w(ij)kl_IsotropicConstraints_Local}
\end{align}

Next, we insert \eqref{tWijkl_Isotropic} and \eqref{tW(ij)kl_IsotropicA} into  \eqref{tW(ijkl)VstWijkl} to obtain 
\begin{align}
\AFive=\AFour
\label{w(ij)kl_IsotropicConstraints_Local_Zero}
\end{align}
whose derivation will be given  in Subsection \ref{Subsec:RelationshipBetween}.
 Consequently, \eqref{tW(ij)kl_IsotropicA} reduces to
\begin{align}
&
\tW_{(IJ)KL}(\bk,\bl)
\notag\\[4pt]
=\,&
\Delta_{Kk}(\bk)\,
\Delta_{Ll}(\bl)
\Big[
 \delta_{IJ}\,\delta_{kl}\,\AOne(k,l,|\bk+\bl|)
+(\delta_{Ik}\,\delta_{Jl}+\delta_{Il}\,\delta_{Jk})\,\ATwo(k,l,|\bk+\bl|)
\notag\\[4pt]&\hskip27mm
+\delta_{IJ}\,l_k\,k_l\,\AThree(k,l,|\bk+\bl|)
+(k_I+l_I)\,(k_J+l_J)\,l_k\,k_l\,\AFour(k,l,|\bk+\bl|)
\Big]
\label{tW(ij)kl_Isotropic}
\end{align}

We further constrain $\Ai$ by imposing the two integral constraints of \eqref{Homogeneity_Symmetry_Inversion_fs}, 
\begin{align*}
&
\int_{\mathbb{R}^3} d\bm
 \l[\tW_{(IJ)KL}(\bm+\bk,-\bm)-\tW_{(KL)IJ}(\bm+\bk,-\bm)\r]=0,
\notag\\& 
\int_{\mathbb{R}^3} d\bm
 \l[\tW_{(IJ)KL}(\bm,\bk)-\tW_{(IK)JL}(\bm,\bk)\r]=0
\end{align*}
Considering the tensor character of these relations,
we analyze them in the special coordinate system where $\bk=(0,0,k)$, under  $\bk\not=\bo$, 
(the case of $\bk=\bo$ is accounted for through continuity).
Due to the adoption of this special coordinate system, many integrands involved in the above relations
are odd functions of  $m_1$ or $m_2$ or are invariant under the interchange between $m_1$ and $m_2$, 
and thus, the associated integrals are trivial.
With a lengthy but straight-forward procedure to evaluate the above two relations component-wise, we get
\begin{align}
&
\int_0^{+\infty}dm\,\int_{|m-k|}^{m+k} d|\bm+\bk|\,
\frac{m}{|\bm+\bk|}
\notag\\[4pt]&\hskip10mm\times
\Big[
\Big(
\big(
|\bm+\bk|^2
-2\,k^2
-3\,km\Theta
\big)\big(1-\Theta^2\big)
-2\,|\bm+\bk|^2\,\Theta^2
\Big)\,\AOne(|\bk+\bm|,m,k)
\notag\\[4pt]&\hskip18mm
+\Big(
 2\,|\bm+\bk|^2
-k^2
-3\,m^2\,\Theta^2
-4\,km\Theta
\Big) \big(1-\Theta^2\big)\,k^2\,\AThree(|\bk+\bm|,m,k)
\notag\\[4pt]&\hskip18mm
-km\Theta\,\big(km\Theta+k^2\big)\,\big(1-\Theta^2\big)\,k^2\,
                   \AFour(|\bk+\bm|,m,k)
\Big]
=0,
\notag\\[4pt]&
\int_0^{+\infty}dm\,\int_{|m-k|}^{m+k} d|\bm+\bk|\,m\,
  \frac{|\bm+\bk|^2-m^2}{|\bm+\bk|}\,\big(1-\Theta^2\big)\,\ATwo(|\bk+\bm|,m,k)=0,
\notag\\[4pt]&
\int_0^{+\infty}dm\,\int_{|m-k|}^{m+k} d|\bm+\bk|\,m\,|\bm+\bk|
\Big[
 \big(1+\Theta^2\big) \Big(\AOne(m,k,|\bm+\bk|)-\ATwo(m,k,|\bm+\bk|)\Big)
\notag\\[4pt]&\hskip67mm
-km\Theta\,\big(1-\Theta^2\big)\,\AThree(m,k,|\bm+\bk|)
\Big]=0,
\notag\\[4pt]&
\int_0^{+\infty}dm\,\int_{|m-k|}^{m+k} d|\bm+\bk|\,m\,|\bm+\bk|
\notag\\[4pt]&\hskip10mm\times
\Big[
 \big(1+\Theta^2\big)\,\AOne(m,k,|\bm+\bk|)
-2\,\big(1-\Theta^2\big)\,\ATwo(m,k,|\bm+\bk|)
\notag\\[4pt]&\hskip18mm
-km\Theta\,\big(1-\Theta^2\big)\,\AThree(m,k,|\bm+\bk|)
\notag\\[4pt]&\hskip18mm
-\big(km\Theta+m^2\big)\,\big(km\Theta+k^2\big)\,\big(1-\Theta^2\big)\,
     \AFour(m,k,|\bm+\bk|)
\Big]=0,
\notag\\&
 \Theta=\frac{|\bm+\bk|^2-m^2-k^2}{2\,k\,m}
\label{w(ij)kl_IsotropicConstraints_Global}
\end{align}
In these integral constraints,
we adopt the conventional spherical coordinate system to 
implement $\int_{\mathbb{R}^3}d\bm$
due to its convenience, 
\begin{align}
&
m_1=m\,\sin\theta\,\cos\phi,\quad
m_2=m\,\sin\theta\,\sin\phi,\quad
m_3=m\,\cos\theta,
\notag\\[4pt]&
d\bm=m^2\,\sin\theta\,dm\,d\phi\,d\theta,\quad
 m\in(0,+\infty),\quad \phi\in[0,2\pi),\quad \theta\in[0,\pi]
\label{SphericalCoordinateSystem}
\end{align}
followed by the change of variables $\Theta=\cos\theta$
and $\Theta\rightarrow |\bm+\bk|$ 
in order to match with the arguments of the scalar functions involved.

\subsection{Relationship Between General and  Degenerated Fourth Order
Correlations}\label{Subsec:RelationshipBetween}
\ \ \ \
In this subsection, we give a rather detailed analysis on
 what invariants should be involved in $\AAOne(\bk,\bl,\bm)$,
 why \eqref{w(ij)kl_IsotropicConstraints_Local_Zero} needs to hold,
 and what possibly additional constraints are present for $\AAOne$,
by exploiting fully \eqref{tW(ijkl)VstWijkl}.
This analysis also helps to simplify the treatment of \eqref{HIT_CLMInPhysicalSpace_www_fs}
 if one intends to solve for $\tW_{IJKL}$ and $\AAOne$,
since \eqref{HIT_CLMInPhysicalSpace_www_fs} involves directly $\tW_{(IJ)KL}$ 
 which acts as an intermediate variable.

Substituting \eqref{tWijkl_Isotropic} and \eqref{tW(ij)kl_IsotropicA} into  \eqref{tW(ijkl)VstWijkl},
we obtain
\begin{align}
&
\Delta_{Kk}(\bk)\,
\Delta_{Ll}(\bl)
\Big[
 \delta_{IJ}\,\delta_{kl}\,\AOne(k^2,l^2,|\bk+\bl|^2)
+(\delta_{Ik}\,\delta_{Jl}+\delta_{Il}\,\delta_{Jk})\,\ATwo(k^2,l^2,|\bk+\bl|^2)
\notag\\[4pt]&\hskip27mm
+\delta_{IJ}\,(k_k+l_k)\,(k_l+l_l)\,\AThree(k^2,l^2,|\bk+\bl|^2)
\notag\\[4pt]&\hskip27mm
+(l_I\,k_J+k_I\,l_J)\,(k_k+l_k)\,(k_l+l_l)\,\AFour(k^2,l^2,|\bk+\bl|^2)
\notag\\[4pt]&\hskip27mm
+k_I\,k_J\,(k_k+l_k)\,(k_l+l_l)\,\AFive(k^2,l^2,|\bk+\bl|^2)
\notag\\[4pt]&\hskip27mm
+l_I\,l_J\,(k_k+l_k)\,(k_l+l_l)\,\AFive(l^2,k^2,|\bk+\bl|^2)
\Big]
\notag\\[4pt]
=\,&
\Delta_{Kk}(\bk)\,
\Delta_{Ll}(\bl)
\int_{\mathbb{R}^3} d\bm\,\Delta_{Ii}(\bn)\,\Delta_{Jj}(\bm)\,
\Big(
 \delta_{ij}\,\delta_{kl}\,\AAOne(\bm,\bk,\bl)
+\delta_{ik}\,\delta_{jl}\,\AAOne(\bk,\bm,\bl)
\notag\\[4pt]&\hskip69mm
+\delta_{il}\,\delta_{jk}\,\AAOne(\bl,\bk,\bm)
\Big),
\ \ \bn=\bm+\bk+\bl
\label{wwwwScalarLink_IJkl}
\end{align}
Here, we have adopted an equivalent set of (squared) arguments for $\Ai$
for the sake of convenience in discussion.
Equality \eqref{wwwwScalarLink_IJkl} is a tensor equation, 
and because of this tensor character, we analyze the
equation in the special coordinate system where  
\begin{align}
 k_1=l_1=0, \quad
\bk+\bl=(0,0,|\bk+\bl|)
\label{wwwwScalarLink_IJkl_SpecialCoordinateSystem}
\end{align}
This is the case of the concerned $\bk$ and $\bl$ forming a plane, and the coordinate system is rotated
such that \eqref{wwwwScalarLink_IJkl_SpecialCoordinateSystem} holds in the rotated  system;
The exceptional case of $\bk$ and $\bl$ parallel to each other or zero can be dealt with 
through the continuous distributions of the scalar functions and a limiting procedure.
This special coordinate system helps to simplify  the mathematical operations and motivates us
to adopt the form of $(k_k+l_k)(k_l+l_l)$ in the above equation.
One consequence of \eqref{wwwwScalarLink_IJkl_SpecialCoordinateSystem}
 is that both $|\bm+\bk+\bl|$ and $\AAOne(\bm,\bk,\bl)$ are even functions of $m_1$,
which makes many integrals involved in \eqref{wwwwScalarLink_IJkl} automatically zero.

The interrelationship \eqref{wwwwScalarLink_IJkl} needs to be evaluated component-wise. 
Firstly,
the components of $IJKL=2311$, $3211$ result in
\begin{align}
&
\int_{\mathbb{R}^3} d\bm\,\Delta_{2j}(\bn)\,\Delta_{3j}(\bm)\,\AAOne(\bm,\bk,\bl)
+\int_{\mathbb{R}^3} d\bm\,\Delta_{12}(\bn)\,\Delta_{13}(\bm)\,\Big[\AAOne(\bk,\bm,\bl)+\AAOne(\bl,\bk,\bm)\Big]
=0,
\notag\\[4pt]&
 \int_{\mathbb{R}^3} d\bm\,\Delta_{3j}(\bn)\,\Delta_{2j}(\bm)\,\AAOne(\bm,\bk,\bl)
+\int_{\mathbb{R}^3} d\bm\,\Delta_{13}(\bn)\,\Delta_{12}(\bm)\,\Big[\AAOne(\bk,\bm,\bl)+\AAOne(\bl,\bk,\bm)\Big]
=0,
\notag\\[4pt]& \bn=\bm+\bk+\bl
\label{wwwwScalarLink_Type1}
\end{align}
These two integral equalities supposedly contain only the invariants,
$k$, $l$ and $|\bk+\bl|$, reflecting the nature of isotropic turbulence.
We have assumed that $\AAOne$ depends possibly on the invariants of $\bk$, $\bl$ and $\bm$,
\begin{align}
\AAOne(\bk,\bl,\bm)=\hat\AAOne\big(k^2, l^2, m^2, |\bk+\bl|^2, |\bl+\bm|^2, |\bm+\bk|^2\big)
\label{tWijkl_Isotropic_Coefficient_Intermediate}
\end{align}
 with
\begin{align}
  |\bl+\bm|^2=l^2+m^2-2(k_2\,m_2+k_3\,m_3-|\bk+\bl|\,m_3),\ \
 |\bm+\bk|^2=k^2+m^2+2(k_2\,m_2+k_3\,m_3)
\end{align}
To guarantee the non-explicit presence of components $k_2$ and $k_3$ 
in \eqref{wwwwScalarLink_Type1},
we need to combine the above two quantities so as to eliminate the components,
\begin{align}
 |\bl+\bm|^2+|\bm+\bk|^2= k^2+l^2+2\,m^2+2\,|\bk+\bl|\,m_3
\end{align}
which in turn can be equivalently replaced by 
\begin{align}
 |\bk+\bl+\bm|^2=|\bk+\bl|^2+m^2+2\,|\bk+\bl|\,m_3
\end{align}
along with $k^2$, $l^2$ and $m^2$.
Next, we need to exclude the presence of $|\bk+\bl|^2$ as an independent argument in $\hat\AAOne$ of
 \eqref{tWijkl_Isotropic_Coefficient_Intermediate}
 in order to satisfy the symmetry property \eqref{tWijkl_Isotropic_Coefficient}
and to guarantee no separate presence of $|\bl+\bm|^2$ and $|\bm+\bk|^2$
in the integrands of \eqref{wwwwScalarLink_Type1}. 
 Otherwise, such separate presences would result from the combined effect of
  both $|\bk+\bl|^2$ as one of the invariant arguments
  of \eqref{tWijkl_Isotropic_Coefficient_Intermediate}
 and the various interchanged positions of $\bk$, $\bl$ and $\bm$ in $\AAOne$ in  \eqref{wwwwScalarLink_Type1}.
Thus, to have the desired invariants, $k$, $l$ and $|\bk+\bl|$, present in \eqref{wwwwScalarLink_Type1},
 $\AAOne$ depends possibly on the invariants as follows, 
\begin{align}
\AAOne(\bm,\bk,\bl)=\AAOne\big(|\bm+\bk+\bl|^2, m^2, k^2, l^2\big)
\label{tWijkl_Isotropic_Coefficient_InvariantsDependent}
\end{align}
constrained by 
\begin{align}
&
 \AAOne\big(|\bk+\bl+\bm|^2,k^2, l^2, m^2 \big)
=\AAOne\big(|\bk+\bl+\bm|^2,k^2, m^2, l^2\big)
=\AAOne\big(k^2,|\bk+\bl+\bm|^2,l^2,m^2\big)
\notag\\[4pt]&
=\AAOne\big(l^2,m^2, |\bk+\bl+\bm|^2, k^2 \big)
\label{tWijkl_Isotropic_CoefficientReduced}
\end{align}
following from \eqref{tWijkl_Isotropic_Coefficient}.

Under \eqref{wwwwScalarLink_IJkl_SpecialCoordinateSystem} and \eqref{tWijkl_Isotropic_Coefficient_InvariantsDependent},
we have the property that
\begin{align}
 \text{$|\bm+\bk+\bl|$ and $\AAOne(\bm,\bk,\bl)$ are even functions of $m_1$ and $m_2$ 
 and invariant under $m_1\leftrightarrow m_2$}
\label{StrongSymmetryProperty}
\end{align}
This property guarantees the automatic satisfaction of \eqref{wwwwScalarLink_Type1}
and is used to simplify the constraints of other components below.

Secondly,
the components of $IJ=12,13,21,31$ and the symmetry properties of $\AAOne$ and $\ATwo$, 
\eqref{tWijkl_Isotropic_Coefficient} and \eqref{w(ij)kl_IsotropicConstraints_Local}, yield
\begin{align}
\ATwo(k^2,l^2,|\bk+\bl|^2)
=\,&
 \int_{\mathbb{R}^3}d\bm\,
\Big(\Delta_{11}(\bn)\,\Delta_{22}(\bm)+\Delta_{12}(\bn)\,\Delta_{12}(\bm)\Big) \AAOne(\bk,\bm,\bl)
\notag\\[4pt]
=\,&
 \int_{\mathbb{R}^3} d\bm\,
\Big(\Delta_{11}(\bn)\,\Delta_{33}(\bm)+\Delta_{13}(\bn)\,\Delta_{13}(\bm)\Big) \AAOne(\bk,\bm,\bl)
\notag\\[4pt]
=\,&
 \int_{\mathbb{R}^3} d\bm\,
 \Big(\Delta_{11}(\bn)\,\Delta_{33}(\bm)+\Delta_{13}(\bn)\,\Delta_{13}(\bm)\Big)\,\AAOne(\bl,\bm,\bk)
,
\ \ \bn=\bm+\bk+\bl
\label{wwwwScalar1ink_TypeII}
\end{align}
where the integrations are implemented with the help of \eqref{SphericalCoordinateSystem}.

Thirdly, 
the components of $IJKL=1111,2211,3311$ result in
\begin{align}
&
\AOne(k^2,l^2,|\bk+\bl|^2)
\notag\\[4pt]
=\,&
 \int_{\mathbb{R}^3} d\bm\,\Delta_{1j}(\bn)\,\Delta_{1j}(\bm)\,\AAOne(\bm,\bk,\bl)
+2\int_{\mathbb{R}^3} d\bm\,\Delta_{12}(\bn)\,\Delta_{12}(\bm)\,\AAOne(\bk,\bm,\bl)
\notag\\[4pt]
=\,&
 \int_{\mathbb{R}^3} d\bm\,\Delta_{j3}(\bn)\,\Delta_{j3}(\bm)\,\AAOne(\bm,\bk,\bl)
+2\int_{\mathbb{R}^3} d\bm\,\Delta_{13}(\bn)\,\Delta_{13}(\bm)\,\AAOne(\bk,\bm,\bl)
\notag\\[4pt]
=\,&
 \int_{\mathbb{R}^3} d\bm\,\Delta_{1j}(\bn)\,\Delta_{1j}(\bm)\,\AAOne(\bm,\bk,\bl)
+2\int_{\mathbb{R}^3} d\bm\,\Delta_{11}(\bn)\,\Delta_{11}(\bm)\,\AAOne(\bk,\bm,\bl)
\notag\\[4pt]&
-2\,\ATwo(k^2,l^2,|\bk+\bl|^2),\ \  \bn=\bm+\bk+\bl
\label{wwwwScalarLinkUpdated_(11,22,33)11_TypeIII}
\end{align}
The last equality indicates that these expressions for $\AOne$ can also be cast as 
the expressions for $\ATwo$.

Fourthly,
the component of $IJKL=1122$ leads to
\begin{align}
&
\AThree(k^2,l^2,|\bk+\bl|^2)
=
 \frac{2}{|\bk+\bl|^2}\int_{\mathbb{R}^3} d\bm\,\Big(\Delta_{13}(\bn)\,\Delta_{13}(\bm)-\Delta_{12}(\bn)\,\Delta_{12}(\bm)\Big)
               \AAOne(\bk,\bm,\bl),
\notag\\[4pt]& \bn=\bm+\bk+\bl
\label{wwwwScalarLinkUpdated_1122_TypeIV}
\end{align}
which  provides the integral relationship between $\AThree$ and $\AAOne$.

Fifthly, we evaluate the equalities under $IJKL=2222$, $2322$ and employ the linear independent properties
 of  $\{1, (k_2)^2\}$ and  $\{1,k_3,(k_3)^2\}$ to obtain
\begin{align}
\AFour=\AFive
\label{wwwwScalarLinkUpdated_2322_TypeV}
\end{align}

Finally, the equality from $IJKL=3322$ gives
\begin{align}
&
 2\,\ATwo(k^2,l^2,|\bk+\bl|^2)
+|\bk+\bl|^2\,\AThree(k^2,l^2,|\bk+\bl|^2)
+|\bk+\bl|^4\,\AFour(k^2,l^2,|\bk+\bl|^2)
\notag\\[4pt]
=\,&
 2\int_{\mathbb{R}^3} d\bm\,\Big(\Delta_{33}(\bn)\,
  \Delta_{33}(\bm)-\Delta_{13}(\bn)\,\Delta_{13}(\bm)\Big)\AAOne(\bk,\bm,\bl)
,
\ \ \bn=\bm+\bk+\bl
\label{wwwwScalar2inkUpdated_3322_TypeVI}
\end{align}
which provides the integral relationship between $\AFour$ and $\AAOne$.

A few comments are worth to make here.
The multiple expressions for $\ATwo$ in \eqref{wwwwScalar1ink_TypeII}
and for $\AOne$ in  \eqref{wwwwScalarLinkUpdated_(11,22,33)11_TypeIII}
represent a set of the integral constraints of equality for
the scalar function $\AAOne$.
Based on the role played by $\tW_{(IJ)KL}$ as the intermediate variable,
these constraints may also be expectedly derived from the substitution of 
 \eqref{tW(ijkl)VstWijkl} and \eqref{tWijkl_Isotropic}
 into the tensor equation of evolution \eqref{HIT_CLMInPhysicalSpace_www_fs};
 the component-wise satisfaction
 and implementation of the latter produces these constraints
 and the dynamical equations \eqref{G1Evolution} and \eqref{G4Evolution} below.
 These constraints need to be imposed explicitly,
 if $\tW_{IJKL}$ (or $\AAOne$) is adopted as
 the control variable in the fourth-order model.
 This adoption poses a great challenge computationally -- how to handle adequately
  $\AAOne(\bk,\bl,\bm)=\AAOne(|\bk+\bl+\bm|^2,k^2,l^2,m^2)$
  which results in a larger number of discretized
control variables and constraints,
 the integrals of higher dimensions, etc.
Additionally, there is a subtle issue in the treatment of $\tW_{IJKL}$ (or $\AAOne$)
as a control variable,
since the evolution equations, \eqref{HIT_PressureInPhysicalSpace_qq_fs} through
\eqref{HIT_CLMInPhysicalSpace_www_fs},
involve only $\tW_{(IJ)KL}$ and additional information contained in the solution
of $\tW_{IJKL}$
 expectedly comes from the related constraints and optimization.
Next, the required satisfaction of the constraints for $\AAOne$ 
indicates the difficulty to approximate $\tW_{IJKL}$ appropriately
 in terms of lower order  correlations as done in  a conventional closure scheme,
because of the great difference between the information contents 
 in these correlations of various orders and various numbers of vectors involved.

\subsection{Primary Dynamical Equations of Evolution}
\ \ \ \
With the help of the isotropic forms presented in \eqref{tQj_Isotropic} through
 \eqref{tWijk_Isotropic} and \eqref{tW(ij)kl_Isotropic} and
 with the help of the special coordinate system in which either $\bk=(0,0,k)$ or
 \eqref{wwwwScalarLink_IJkl_SpecialCoordinateSystem} holds,
we can derive from \eqref{HIT_CLMInPhysicalSpace_ww_fs} and \eqref{HIT_CLMInPhysicalSpace_www_fs} 
the following dynamical equations governing the evolution of the scalar functions $\tW_{kk}$, $\GOne$ and $\GFour$,
\begin{align}
&
 \bigg(\frac{\partial}{\partial t}+2\,\nu\,k^2\bigg)\tW_{kk}(k)   
\notag\\[4pt]
=\,&
4\pi \int_0^{+\infty}dl \int_{|l-k|}^{l+k}d|\bk+\bl|\,l\,|\bk+\bl|\,
       \big(1-\Theta^2\big)\,
\notag\\[4pt]&\hskip12mm\times
\bigg[
 \frac{l^2\,(k+l\,\Theta)}{|\bk+\bl|^2}\,\big(1-\Theta^2\big)\,k^2\, 
 \GOne(|\bk+\bl|,k,l)
+\bigg(
 \Theta
+\frac{k\,l\,(1-\Theta^2)}{|\bk+\bl|^2}
   \bigg)\,l\,\GFour(|\bk+\bl|,k,l)
\notag\\[4pt]&\hskip20mm
-\bigg(
       2
      +\frac{2\,l^2\,\Theta^2+k\,l\,\Theta-l^2}{|\bk+\bl|^2}
   \bigg)\,k\,\GFour(|\bk+\bl|,l,k)
\bigg],
\ \ 
\Theta=\frac{|\bk+\bl|^2-k^2-l^2}{2\,k\,l}
\label{twkkEvolution}
\end{align}
\begin{align}
 \bigg(\frac{\partial }{\partial t}+\nu\big(k^2+l^2+|\bk+\bl|^2\big)\bigg)
  \GOne(|\bk+\bl|,k,l)
=0 
\label{G1Evolution}
\end{align}
and
\begin{align}
 \bigg(\frac{\partial }{\partial t}+\nu\big(k^2+l^2+|\bk+\bl|^2\big)\bigg)\GFour(l,|\bk+\bl|,k)
= \ATwo(|\bk+\bl|,k,l)-\ATwo(|\bk+\bl|,l,k)
\label{G4Evolution}
\end{align}

These dynamical equations have several interesting features.
The first is that equation \eqref{twkkEvolution} is another version of the known
 (see Equations (52) and (54)  of \cite{ProudmanReid1954});
 the change of variables, $\Theta\rightarrow |\bk+\bl|$,
 is adopted to be suitable to the arguments of the scalar functions involved.
The second is that the evolution of $\GOne$ has a zero source,
 the same as Equation (59) of \cite{ProudmanReid1954}.
The third is that the evolution of $\GFour$ is directly affected only by $\ATwo$, 
the impacts of $\AOne$, $\AThree$ and $\AFour$ are
through the integral constraints of
\eqref{w(ij)kl_IsotropicConstraints_Global}
 and the constraints to be formulated.

With the help of \eqref{w(ij)kl_IsotropicConstraints_Local}, we can infer from \eqref{G4Evolution} that
\begin{align}
 \bigg(\frac{\partial }{\partial t}+\nu\big(k^2+l^2+|\bk+\bl|^2\big)\s\s\bigg)\s
\Big(\GFour(l,|\bk+\bl|,k)+\GFour(|\bk+\bl|,k,l)+\GFour(k,l,|\bk+\bl|)\s\Big)
=0
\label{SummationG4Evolution}
\end{align}
which is equivalent to Equation (62) of \cite{ProudmanReid1954}.
Both \eqref{G1Evolution} and \eqref{SummationG4Evolution} 
may be useful in the analysis of asymptotic state solutions at large time.

\subsection{Constraints}\label{subsec:Constraints}
\ \ \ \
Within the fourth-order model with $\tW_{(ij)KL}$, or equivalently $\Ai$, $i=1,2,3,4$, 
as the control variables, the following sets of constraints need to be satisfied.

Firstly, we have the constraints of equality \eqref{w(ij)kl_IsotropicConstraints_Local} and  \eqref{w(ij)kl_IsotropicConstraints_Global},
 which involve only the scalar control variables. 
These constraints are to guarantee the self-consistency of the definition of the degenerated fourth 
order correlation introduced in \eqref{Homogeneity}.

Secondly, to satisfy the constraints of equality \eqref{tWijk_Isotropic_Constraints}, 
we need to impose them on the initial conditions of $\GOne$ and $\GFour$, 
as required and guaranteed by \eqref{G1Evolution} and \eqref{G4Evolution}.
The equality of \eqref{tQj_Isotropic} is satisfied automatically,
 which can be directly verified under $\bk=(0,0,k)$.

Thirdly, the non-negativity of the energy spectrum $k^2\,\tW_{kk}(k)/2$ requires that
\begin{align}
 \tW_{kk}(k)\geq 0
\label{twkk_NonNegativeConstraint}
\end{align}
This inequality has several consequences similar to those discussed in PART III \cite{Tao2015a}.
It results in the positive semi-definiteness of the second order vorticity correlation
 $\overline{\tilde\vort_i\tilde\vort_{j}}(\bk)$ defined through
\begin{align}
&
\vort_{j}(\bx)=\epsilon_{jkl}\,\w_{k,l}(\bx),\quad 
\overline{\vort_i(\bx) \vort_{j}(\by)}
=\int_{\mathbb{R}^3}d\bk\, \overline{\tilde\vort_i\tilde\vort_{j}}(\bk)\,\cos(\bk\cdot\br),
\quad
\overline{\tilde\vort_i\tilde\vort_{j}}(\bk)=k^2\,\tW_{ij}(\bk)
\end{align}
It leads to the following inequality for $\overline{\tw_i\tilde\vort_{j}}(\bk)$,
\begin{align}
&
\Big|\overline{\tw_{i}\tilde\vort_{j}}(\bk)\Big|^2
\leq
\tW_{\underlinei\underlinei}(\bk)\,\overline{\tilde\vort_{\underlinej}\tilde\vort_{\underlinej}}(\bk),\quad
\overline{\w_i(\bx) \vort_{j}(\by)}
=\imaginary\int_{\mathbb{R}^3}d\bk\, \overline{\tw_i\tilde\vort_{j}}(\bk)\,\sin(\bk\cdot\br),
\notag\\[4pt]&
\overline{\tw_i\tilde\vort_{j}}(\bk)= \frac{\imaginary}{2}\,\epsilon_{jil}\,k_l\,\tW_{kk}(k)
\end{align}
Furthermore, the corresponding inequalities held automatically in the physical space can be 
established, as done in \cite{Tao2015a},
\begin{align}
\W_{\underlinei\,\underlinei}(\bo)\geq 0,\quad 
\Big|\W_{ij}(\br)\Big|^2
\leq
\W_{\underlinei\,\underlinei}(\bo)\,\W_{\underlinej\,\underlinej}(\bo),\quad
\Big|\W_{ij}(\br)\Big|\leq\frac{1}{3}\,\W_{kk}(\bo)
\,\delta_{\underlinei\,\underlinei}\,\delta_{\underlinej\,\underlinej}
\label{CS_Inequality_wiwj_PhysicalSpace}
\end{align}
\begin{align}
\overline{\Vorticity_{\underlinei}\Vorticity_{\underlinei}}(\bo)\geq 0,\quad
\Big(\overline{\Vorticity_i\Vorticity_j}(\br)\Big)^2
\leq
\overline{\Vorticity_{\underlinei}\Vorticity_{\underlinei}}(\bo)
\,\,\overline{\Vorticity_{\underlinej}\Vorticity_{\underlinej}}(\bo),\quad
 \overline{\vort_i\vort_{j}}(\br)
:=-\epsilon_{imn}\,\epsilon_{jkl}\,\frac{\partial^2\W_{mk}(\br)}{\partial r_n\partial r_l}
\label{CS_Inequality_Vorticity_PhysicalSpace_a}
\end{align} 
\begin{align}
\Big(\overline{\w_i \Vorticity_j}(\br)\Big)^2
\leq
\W_{\underlinei\,\underlinei}(\bo)
\,\,\overline{\Vorticity_{\underlinej} \Vorticity_{\underlinej}}(\bo),\quad
 \overline{\w_i\vort_{j}}(\br):=\epsilon_{jkl}\,\frac{\partial\W_{ik}(\br)}{\partial r_l}
\label{CS_Inequality_VelocityVorticity_PhysicalSpace_a}
\end{align}
which have the characteristic of spatial degeneracy on the right-hand sides of the inequalities.

Fourthly, besides the above resultant and redundant inequalities,
we can also generate the constraints of inequality involving only the second
order correlation by applying the Cauchy-Schwarz inequality,
 $\big|\overline{a b}\big|^2\leq \overline{a a}\,\overline{b b}$,
to structure functions \cite{Davidson2004} involving only 
  the second order correlation such as
\begin{align}
&
\overline{[\w_i(\by)-\w_i(\bx)+\alpha (\w_i(\bz')-\w_i(\bz))]\,
           [\w_j(\bz')-\w_j(\bz)-\beta(\w_j(\by)-\w_j(\bx))]},
\notag\\[4pt]&
\overline{[\w_i(\by)-\w_i(\bx)+\alpha (\w_i(\bz')-\w_i(\bz))]\,
      [\Vorticity_j(\bz')-\Vorticity_j(\bz)-\beta(\Vorticity_j(\by)-\Vorticity_j(\bx))]},
\notag\\[4pt]&
\overline{[\Vorticity_i(\by)-\Vorticity_i(\bx)+\alpha (\Vorticity_i(\bz')-\Vorticity_i(\bz))]\,
      [\Vorticity_j(\bz')-\Vorticity_j(\bz)-\beta(\Vorticity_j(\by)-\Vorticity_j(\bx))]},\ \
      \alpha\geq 0,\ \beta\geq 0
\label{StructureFunctions_SOC}
\end{align}
which results in the following primary constraints of inequality,
\begin{align}
&
\Big[
 \W_{ij}(\bs'-\br)
-\W_{ij}(\bs-\br)
-\W_{ij}(\bs')
+\W_{ij}(\bs)
+2\,\alpha \Big(\W_{ij}(\bo)-\W_{ij}(\bs'-\bs)\Big)
 \notag\\[4pt]&\hskip5mm
-2\,\beta \Big(\W_{ij}(\bo)-\W_{ij}(\br)\Big)
-\alpha\beta \Big(
 \W_{ij}(\bs'-\br)
-\W_{ij}(\bs-\br)
-\W_{ij}(\bs')
+\W_{ij}(\bs)
 \Big)
 \Big]^2
\notag\\[4pt]
\leq\,&
4\,\Big[
 \W_{\underlinei\,\underlinei}(\bo)
-\W_{\underlinei\,\underlinei}(\br)
+\alpha^2 \Big(
 \W_{\underlinei\,\underlinei}(\bo)
-\W_{\underlinei\,\underlinei}(\bs'-\bs)
 \Big)
 \notag\\[4pt]&\hskip10mm
+\alpha \Big(
 \W_{\underlinei\,\underlinei}(\bs'-\br)
-\W_{\underlinei\,\underlinei}(\bs-\br)
-\W_{\underlinei\,\underlinei}(\bs')
+\W_{\underlinei\,\underlinei}(\bs)
\Big)
\Big]
\notag\\[4pt]&\times
\Big[
 \W_{\underlinej\,\underlinej}(\bo)
-\W_{\underlinej\,\underlinej}(\bs'-\bs)
+\beta^2 \Big(
 \W_{\underlinej\,\underlinej}(\bo)
-\W_{\underlinej\,\underlinej}(\br)
 \Big)
 \notag\\[4pt]&\hskip10mm
-\beta \Big(
 \W_{\underlinej\,\underlinej}(\bs'-\br)
-\W_{\underlinej\,\underlinej}(\bs-\br)
-\W_{\underlinej\,\underlinej}(\bs')
+\W_{\underlinej\,\underlinej}(\bs)
\Big)
\Big]
   ,\ \ ij=11,12
\label{StructureFunction_wwIneq_SOC}
\end{align}
\begin{align}
&
\Big[
 \overline{\w_i\Vorticity_j}(\bs'-\br)
-\overline{\w_i\Vorticity_j}(\bs-\br)
-\overline{\w_i\Vorticity_j}(\bs')
+\overline{\w_i\Vorticity_j}(\bs)
 \notag\\[4pt]&\hskip5mm
+\alpha\beta \Big(
 \overline{\w_i\Vorticity_j}(\bs'-\br)
-\overline{\w_i\Vorticity_j}(\bs-\br)
-\overline{\w_i\Vorticity_j}(\bs')
+\overline{\w_i\Vorticity_j}(\bs)
 \Big)
 \Big]^2
\notag\\[4pt]
\leq\,&
4\,\Big[
 \W_{\underlinei\,\underlinei}(\bo)
-\W_{\underlinei\,\underlinei}(\br)
+\alpha^2 \Big(
 \W_{\underlinei\,\underlinei}(\bo)
-\W_{\underlinei\,\underlinei}(\bs'-\bs)
 \Big)
 \notag\\[4pt]&\hskip10mm
+\alpha \Big(
 \W_{\underlinei\,\underlinei}(\bs'-\br)
-\W_{\underlinei\,\underlinei}(\bs-\br)
-\W_{\underlinei\,\underlinei}(\bs')
+\W_{\underlinei\,\underlinei}(\bs)
\Big)
\Big]
\notag\\[4pt]&\times
\Big[
 \overline{\Vorticity_{\underlinej}\Vorticity_{\underlinej}}(\bo)
-\overline{\Vorticity_{\underlinej}\Vorticity_{\underlinej}}(\bs'-\bs)
+\beta^2 \Big(
 \overline{\Vorticity_{\underlinej}\Vorticity_{\underlinej}}(\bo)
-\overline{\Vorticity_{\underlinej}\Vorticity_{\underlinej}}(\br)
 \Big)
 \notag\\[4pt]&\hskip10mm
-\beta \Big(
 \overline{\Vorticity_{\underlinej}\Vorticity_{\underlinej}}(\bs'-\br)
-\overline{\Vorticity_{\underlinej}\Vorticity_{\underlinej}}(\bs-\br)
-\overline{\Vorticity_{\underlinej}\Vorticity_{\underlinej}}(\bs')
+\overline{\Vorticity_{\underlinej}\Vorticity_{\underlinej}}(\bs)
\Big)
\Big],\ \ ij=11,12
\label{StructureFunction_wvIneq_SOC}
\end{align}
and
\begin{align}
&
\Big[
 \overline{\Vorticity_i\Vorticity_j}(\bs'-\br)
-\overline{\Vorticity_i\Vorticity_j}(\bs-\br)
-\overline{\Vorticity_i\Vorticity_j}(\bs')
+\overline{\Vorticity_i\Vorticity_j}(\bs)
+2\,\alpha \Big(
 \overline{\Vorticity_i\Vorticity_j}(\bo)
-\overline{\Vorticity_i\Vorticity_j}(\bs'-\bs)
\Big)
 \notag\\[4pt]&\hskip5mm
-2\,\beta \Big(
 \overline{\Vorticity_i\Vorticity_j}(\bo)
-\overline{\Vorticity_i\Vorticity_j}(\br)
 \Big)
-\alpha\beta \Big(
 \overline{\Vorticity_i\Vorticity_j}(\bs'-\br)
-\overline{\Vorticity_i\Vorticity_j}(\bs-\br)
-\overline{\Vorticity_i\Vorticity_j}(\bs')
+\overline{\Vorticity_i\Vorticity_j}(\bs)
 \Big)
 \Big]^2
\notag\\[4pt]
\leq\,&
4\,\Big[
 \overline{\Vorticity_{\underlinei}\Vorticity_{\underlinei}}(\bo)
-\overline{\Vorticity_{\underlinei}\Vorticity_{\underlinei}}(\br)
+\alpha^2 \Big(
 \overline{\Vorticity_{\underlinei}\Vorticity_{\underlinei}}(\bo)
-\overline{\Vorticity_{\underlinei}\Vorticity_{\underlinei}}(\bs'-\bs)
 \Big)
 \notag\\[4pt]&\hskip10mm
+\alpha \Big(
 \overline{\Vorticity_{\underlinei}\Vorticity_{\underlinei}}(\bs'-\br)
-\overline{\Vorticity_{\underlinei}\Vorticity_{\underlinei}}(\bs-\br)
-\overline{\Vorticity_{\underlinei}\Vorticity_{\underlinei}}(\bs')
+\overline{\Vorticity_{\underlinei}\Vorticity_{\underlinei}}(\bs)
\Big)
\Big]
\notag\\[4pt]&\times
\Big[
 \overline{\Vorticity_{\underlinej}\Vorticity_{\underlinej}}(\bo)
-\overline{\Vorticity_{\underlinej}\Vorticity_{\underlinej}}(\bs'-\bs)
+\beta^2 \Big(
 \overline{\Vorticity_{\underlinej}\Vorticity_{\underlinej}}(\bo)
-\overline{\Vorticity_{\underlinej}\Vorticity_{\underlinej}}(\br)
 \Big)
 \notag\\[4pt]&\hskip10mm
-\beta \Big(
 \overline{\Vorticity_{\underlinej}\Vorticity_{\underlinej}}(\bs'-\br)
-\overline{\Vorticity_{\underlinej}\Vorticity_{\underlinej}}(\bs-\br)
-\overline{\Vorticity_{\underlinej}\Vorticity_{\underlinej}}(\bs')
+\overline{\Vorticity_{\underlinej}\Vorticity_{\underlinej}}(\bs)
\Big)
\Big],\ \  ij=11,12
\label{StructureFunction_vvIneq_SOC}
\end{align}
Here, each inequality involves essentially a scalar function defined 
in a high-dimensional space
composed of the components of $\br$, $\bs$, $\bs'$ and their differences and $t$.
Additional inequalities can be produced similarly based on other structure functions.
We expect that these quadratic constraints
play a significant role to constrain directly the structure of $\tW_{kk}$
and indirectly the structures of the control variables.

Fifthly, for the higher order correlations,
we apply the Cauchy-Schwarz inequality to the correlations of
\begin{align}
\overline{\w_i(\bx) \w_j(\by) \w_k(\bz)},\quad
 \overline{\w_i(\bx) \w_j(\bx) \w_K(\by) \w_L(\bz)},\quad
\overline{\q(\bx) \q(\by)},\quad
\overline{q(\bx) \w_i(\by) \w_j(\bz)}
\end{align}
Here, more are to be added like the correlations involving $\w_{k,l}(\bx)$, $\vort_{j}(\bx)$,
and the structure functions involving $w_i(\by)-\w_i(\bx)$, $\Vorticity_i(\bz')-\Vorticity_i(\bz)$, 
etc. In the derivations of the primary constraints of inequality below, we resort to 
\eqref{Homogeneity_Symmetry}
 and the interchangeability between the space vector components $\{r_j,s_j\}$, $j=1$, $2$, $3$, 
 like $\{r_2,s_2\}\leftrightarrow\{r_1,s_1\}$.
\begin{enumerate}
 \item 
For $\overline{\w_i(\bx) \w_j(\by) \w_k(\bz)}$, there is only one independent decomposition,
\begin{align}
 \Big|\W_{ijk}(\br,\bs)\Big|^2
\leq
 \frac{1}{3}\,\delta_{\underlinek\,\underlinek}\,\W_{ll}(\bo)\,
 \W_{(\underlinei\,\underlinei)\,\underlinej\,\underlinej}(\br,\br),\quad
 \W_{(\underlinei\,\underlinei)\,\underlinej\,\underlinej}(\br,\br)\geq 0
\end{align}
which in turn results in the primary constraints as follows,
\begin{align}
&
 \W_{(\underlinei\,\underlinei)\,\underlinej\,\underlinej}(\br,\br)\geq 0, \ \ i=1,\ j=1,2;
\notag\\[4pt]&
 \Big|\W_{ijk}(\br,\bs)\Big|^2
\leq
 \frac{1}{3}\,\W_{ll}(\bo)\,
 \W_{(\underlinei\,\underlinei)\,\underlinej\,\underlinej}(\br,\br),\quad
 ijk=111, 112, 121, 123
\label{www_CSIneq}
\end{align}

\item
In the case of $\overline{\w_i(\bx) \w_j(\bx) \w_K(\by) \w_L(\bz)}$,
 we need to decompose the quantity properly such that the decomposition correlations are resolvable within the fourth-order model,
\begin{align}
&
 \Big|\W_{(ij)KL}(\br,\bs)\Big|^2
\leq
 \W_{(\underlinei\,\underlinei)\,\underlinej\,\underlinej}(\bo,\bo)\,\,
 \W_{(\underlineK\,\underlineK)\,\underlineL\,\underlineL}(\bs-\br,\bs-\br),
\notag\\[4pt]&
 \Big|\W_{(ij)KL}(\br,\bs)\Big|^2
\leq
 \W_{(\underlinei\,\underlinei)\,\underlineK\,\underlineK}(\br,\br)\,\,
 \W_{(\underlinej\,\underlinej)\,\underlineL\,\underlineL}(\bs,\bs)
\end{align}
whose primary constraints are
\begin{align}
&
 \Big|\W_{(ij)KL}(\br,\bs)\Big|^2
\leq
 \W_{(\underlinei\,\underlinei)\,\underlinej\,\underlinej}(\bo,\bo)\,\,
 \W_{(\underlineK\,\underlineK)\,\underlineL\,\underlineL}(\bs-\br,\bs-\br),
\notag\\[4pt]&\hskip20mm
 ijKL=1111, 1112, 1122, 1123, 1211, 1212, 1213, 1233;
\notag\\[4pt]&
 \Big|\W_{(ij)KL}(\br,\bs)\Big|^2
\leq
 \W_{(\underlinei\,\underlinei)\,\underlineK\,\underlineK}(\br,\br)\,\,
 \W_{(\underlinej\,\underlinej)\,\underlineL\,\underlineL}(\bs,\bs),
\notag\\[4pt]&\hskip20mm
 ijKL=1111, 1112, 1122, 1123, 1211, 1212, 1213, 1221, 1223, 1233
\label{wwww_CSIneq}
\end{align}

\item
For $\overline{\q(\bx) \q(\by)}$,
\vskip-8mm
\begin{align}
|\Q(\br)|\leq \Q(\bo),\ \ \Q(\bo)\geq 0
\label{qq_CSIneq}
\end{align}

\item
In the case of $\overline{q(\bx) \w_i(\by) \w_j(\bz)}$, we obtain
\vskip-6mm
\begin{align}
 \Big|\Q_{ij}(\br,\bs)\Big|^2\leq \Q(\bo)\, \, \W_{(\underlinei \underlinei) \underlinej \underlinej}(\bs-\br,\bs-\br),\ \ ij=11,\ 12
\label{qww_CSIneq}
\end{align}
\vskip-6mm
\end{enumerate}

Sixthly, we formulate the constraints of inequality on the basis of the requirement that the variance of products be non-negative,
\vskip-5mm
\begin{align}
 \overline{\big(XY-\overline{XY}\big)^2}\geq 0,\quad
 \overline{XXYY}\geq\big(\overline{XY}\big)^2
\end{align}
\vskip-4mm
\noindent
We apply it to the specific cases of
\vskip-5mm
\begin{align*}
 (X,Y)=
\big(\w_{i}(\bx),\w_{j}(\by)\big),\ 
\big(\w_{i}(\bx), \vort_{j}(\by)\big),\
\big(\vort_{i}(\bx), \vort_{j}(\by)\big)
\end{align*}
\vskip-4mm
\noindent
to generate
\vskip-10pt
\begin{align}
&
\W_{(\underlinei\underlinei)\underlinej\underlinej}(\br,\br)\geq\Big(\W_{ij}(\br)\Big)^2,
\quad
 \overline{\w_{\underlinei}(\bx)\w_{\underlinei}(\bx)\vort_{\underlinej}(\by)\vort_{\underlinej}(\by)}
\geq\Big(\overline{\w_i\vort_{j}(\br)}\Big)^2,
\notag\\[4pt]&
 \overline{\vort_{\underlinei}(\bx)\vort_{\underlinei}(\bx)\vort_{\underlinej}(\by)\vort_{\underlinej}(\by)}
\geq\Big(\overline{\vort_{i}\vort_{j}(\br)}\Big)^2,\ \ 
ij=11, 12 
\label{VarianceInequality}
\end{align}
\vskip-4pt
\noindent
More such inequalities will be produced like setting $X=w_{k,l}(\bx)$,
$w_i(\by)-\w_i(\bx)$, $\Vorticity_j(\bz')-\Vorticity_j(\bz)$, 
and so on.

The constraints of equality and inequality listed above are an integral part of the 
mathematical setup to model incompressible homogeneous isotropic turbulence.
The issue of how to deal with numerous constraints constructed from structure functions and
the issue of redundancy are not addressed;
 this subject needs to be explored due to its importance to the geometry and size of
 the domain of feasible solutions and the computational feasibility.
Each of the above constraints may be rather loose;
 however, their number is substantial and the very many such constraints 
may form a rather tight restriction on $\Ai$.
These constraints are either linear or quadratically convex, when discretized,
 as functions of the discretized control variables $\Ai$, $i=1,2,3,4$,
since \eqref{twkkEvolution} through \eqref{G4Evolution} are of linear structures,
 similar to that argued in \cite{Tao2015a}.

\subsection{Expressions for Correlations in Physical Space}\label{subsec:ExpressionsforCorrelationsin PhysicalSpace}
\ \ \ \
We present here the expressions for the correlations in the physical space
 which are needed for the implementation of the constraints formed 
 in the physical space.

To this end, we substitute \eqref{HIT_PressureInPhysicalSpace_qq_fs}, \eqref{HIT_PressureInPhysicalSpace_qww_fs}, \eqref{tWij_Isotropic}, \eqref{tWijk_Isotropic} and \eqref{tW(ij)kl_Isotropic}
into \eqref{FourierTransform} and then operate on the resultant expressions with some conventional techniques.
The first technique is to use
\begin{align*}
k_j\, \exp(\imaginary\,\bk\s\cdot\s\br)
=\frac{1}{\imaginary}\,\frac{\partial\exp(\imaginary\,\bk\s\cdot\s\br)}{\partial r_j}
\end{align*}
and the like involving $l_j$. 
With the aid of this technique, we can replace all the wave-number components of free indexes in 
the expressions
with the corresponding partial derivatives with respect to $r_j$ or $s_j$. For example, we obtain
\begin{align}
\Wij(\br)
=
 \frac{1}{2}\,\delta_{ij} \s\int_{\mathbb{R}^3}d\bk\,\tW_{kk}(k)\,\exp(\imaginary\, \bk\s\cdot\s\br)
+\frac{1}{2}\,\frac{\partial^2}{\partial r_i \partial r_j}\s
 \int_{\mathbb{R}^3}d\bk\,\frac{1}{k^2}\,\tW_{kk}(k)\,\exp(\imaginary\, \bk\s\cdot\s\br)
\end{align}
The second is the adoption of the spherical coordinate system \eqref{SphericalCoordinateSystem},
  $\bk\rightarrow(k,\phi,\theta)$ and $\bl\rightarrow(l,\phi',\theta')$,
 in order to introduce $r=|\br|$ (and/or $s=|\bs|$) into the expressions 
 and to simplify the integrations analytically.
Specifically, under given $\br$ and $\bs$, 
we transform/rotate $\bk$ and $\bl$ such that $\bk\cdot\br=r\,k_3$ 
and $\bl\cdot\bs=s\,l_3$ (or $\bl\cdot\bk=k\,l_3$ if $\bs$ is absent and $\bl$ present).
The third is to integrate $\theta\, (\theta')$ and $\phi\, (\phi')$ analytically, if possible, such as 
$\int_0^{2\pi}d\phi\int_0^{2\pi}d\phi'\,f(\cos(\phi-\phi'))
=2\pi\int_0^{2\pi}d\phi\,f(\cos\phi)
=4\pi\int_0^{\pi}d\phi\,f(\cos\phi)=4\pi\int_{-\pi/2}^{\pi/2}d\phi\,f(-\sin\phi)$,
and employ the change of variables
$\Theta=\cos\theta$, $\Theta'=\cos\theta'$, etc.

Using the techniques and procedure outlined, 
we can derive from \eqref{FourierTransform} the following
\begin{align}
&
\Q(\br)
\notag\\[4pt]
=\,&
8\pi^2
\int_0^{+\infty}dk\,\int_0^{+\infty}dl\,\int_{-1}^1d\Theta'\,
 \frac{l^2}{k^2}\,
\notag\\[4pt]&\hskip10mm\times\s
\bigg\{
\AOne(|\bk+\bl|,l,k)\,
 k^2 \bigg(
       k^2
      -\frac{(k^2+\bk\cdot\bl)^2}{|\bk+\bl|^2}
      -\frac{(\bk\cdot\bl)^2}{l^2}
      +\frac{(k^2+\bk\cdot\bl)}{|\bk+\bl|^2}\frac{(\bk\cdot\bl)\,(\bk\cdot\bl+l^2)}{l^2}
                 \bigg)
\notag\\[4pt]&\hskip18mm
+\Big(2\,\ATwo(|\bk+\bl|,l,k)+k^2\,\AThree(|\bk+\bl|,l,k)+k^4\,\AFour(|\bk+\bl|,l,
k)\Big)
\notag\\[4pt]&\hskip25mm\times
\bigg(k^2-\frac{(k^2+\bk\cdot\bl)^2}{|\bk+\bl|^2}\bigg)\bigg(k^2-\frac{(\bk\cdot\bl)^2}{l^2}\bigg)
\bigg\}\, \frac{\sin(r k)}{rk},
\notag\\[4pt]&
\bk\cdot\bl=k\,l\,\Theta',\ \ |\bk+\bl|^2=k^2+l^2+2\,\bk\cdot\bl
\label{qq_IsotropicExpression_ps}
\end{align}
\begin{align}
\Wij(\br)
=\,&
\frac{2\pi}{r}\,\bigg\{
 \delta_{ij}
 \int_0^{+\infty}dk\,k\,
     \bigg( \sin(r k)
           -\frac{\sin(rk)}{r^2 k^2}
           +\frac{\cos(rk)}{r k}
\bigg)\,\tW_{kk}(k)
\notag\\&\hskip10mm
-\frac{r_i r_j}{r^4} \int_0^{+\infty}dk\,
           \frac{(r^2 k^2-3) \sin(rk)+3 r k \cos(rk)}{k}\,\tW_{kk}(k)
\bigg\}
\label{wij_IsotropicExpression_ps}
\end{align}
\begin{align}
&
\Q_{KL}(\br,\bs)
\notag\\[4pt]
=\,&
4\pi\s
\int_0^{+\infty}dk\,
\int_0^{+\infty}dl\,
\int_{-1}^1d\Theta\,
\int_{-1}^1d\Theta'\,
\int_{-\pi/2}^{\pi/2}d\phi\,
 \frac{k^2\,l^2}{|\bk+\bl|^2}\,
\notag\\[4pt]&\hskip10mm\times
\bigg\{
 \AOne(k,l,|\bk+\bl|)\, |\bk+\bl|^2
\bigg(
-\delta_{KL}
-\frac{1}{k^2}\,\frac{\partial^2}{\partial r_K\partial r_L}
-\frac{1}{l^2}\,\frac{\partial^2}{\partial s_K\partial s_L}
+\frac{\bk\cdot\bl}{k^2\,l^2}\,\frac{\partial^2}{\partial r_K\partial s_L}
\bigg)
\notag\\[4pt]&\hskip18mm
+\Big(2\ATwo(k,l,|\bk+\bl|)+|\bk+\bl|^2\,\AThree(k,l,|\bk+\bl|)+|\bk+\bl|^4\,\AFour(k,
l,|\bk+\bl|)\Big)
\notag\\[4pt]&\hskip30mm\times
\bigg(\frac{\partial}{\partial s_K}
            -\frac{\bk\cdot\bl}{k^2}\,\frac{\partial}{\partial r_K}\bigg)
     \bigg(\frac{\partial}{\partial r_L}
           -\frac{\bk\cdot\bl}{l^2}\frac{\partial}{\partial s_L}\bigg)\,
\bigg\}\cos(rk \Theta+sl \Theta')
\label{qwij_IsotropicExpression_ps}
\end{align}
\begin{align}
&
\W_{ijk}(\br,\bs)
\notag\\
=\,&
4\pi
\int_0^{+\infty}dk\,
\int_0^{+\infty}dl\,
\int_{-1}^1 d\Theta\,
\int_{-1}^1 d\Theta'\,
\int_{-\pi/2}^{\pi/2}d\phi\,k^2\,l^2\,
\notag\\[4pt]&\hskip10mm\times
\bigg\{
 -\GOne(m,k,l)
\bigg[
 \frac{\partial}{\partial r_i}
-\frac{k^2+\bk\cdot\bl}{|\bk+\bl|^2}\bigg(\frac{\partial}{\partial r_i}+\frac{\partial}{\partial s_i}\bigg)\bigg]
\bigg(\frac{\partial}{\partial s_j}-\frac{\bk\cdot\bl}{k^2}\frac{\partial}{\partial r_j}\bigg)
\bigg(\frac{\partial}{\partial r_k}-\frac{\bk\cdot\bl}{l^2}\frac{\partial}{\partial s_k}\bigg)
\notag\\[4pt]&\hskip18mm
+\GFour(m,l,k)
\bigg[
 \delta_{ij}
+\frac{1}{k^2}\frac{\partial^2}{\partial r_i \partial r_j}
+\frac{1}{|\bk+\bl|^2}
            \bigg(\frac{\partial}{\partial r_i}+\frac{\partial}{\partial s_i}\bigg)
                \bigg(\frac{\partial}{\partial r_j}+\frac{\partial}{\partial s_j}\bigg)
\notag\\&\hskip50mm
-\frac{k^2+\bk\cdot\bl}{|\bk+\bl|^2\,k^2}\bigg(\frac{\partial}{\partial r_i}+\frac{\partial}{\partial s_i}\bigg)\frac{\partial}{\partial r_j}
\bigg]
\bigg(\frac{\partial}{\partial r_k}-\frac{\bk\cdot\bl}{l^2}\frac{\partial}{\partial s_k}\bigg)
\notag\\[4pt]&\hskip18mm
+\GFour(l,m,k)
\bigg[
 \frac{\partial}{\partial r_i}
-\frac{k^2+\bk\cdot\bl}{|\bk+\bl|^2}\bigg(\frac{\partial}{\partial r_i}+\frac{\partial}{\partial s_i}\bigg)\bigg]
\notag\\&\hskip50mm\times
\bigg(
 \delta_{jk}
+\frac{1}{l^2}\frac{\partial^2}{\partial s_j \partial s_k}
+\frac{1}{k^2}\frac{\partial^2}{\partial r_j \partial r_k}
-\frac{\bk\cdot\bl}{k^2\,l^2}\frac{\partial^2}{\partial r_j \partial s_k}
\bigg)
\notag\\[4pt]&\hskip18mm
+\GFour(m,k,l)
\bigg[
 \delta_{ki}
+\frac{1}{l^2}\frac{\partial^2}{\partial s_k \partial s_i}
+\frac{1}{|\bk+\bl|^2}
     \bigg(\frac{\partial}{\partial r_k}+\frac{\partial}{\partial s_k}\bigg)
          \bigg(\frac{\partial}{\partial r_i}+\frac{\partial}{\partial s_i}\bigg)
\notag\\&\hskip50mm
-\frac{\bk\cdot\bl+l^2}{|\bk+\bl|^2\,l^2}
     \frac{\partial}{\partial s_k}
         \bigg(\frac{\partial}{\partial r_i}+\frac{\partial}{\partial s_i}\bigg)
\bigg]
\bigg(\frac{\partial}{\partial s_j}-\frac{\bk\cdot\bl}{k^2}\frac{\partial}{\partial r_j}\bigg)
\bigg\}\,
\notag\\[4pt]&\hskip10mm\times
\cos(r k \Theta+s l \Theta')
\label{wijk_IsotropicExpression_ps}
\end{align}
and
\begin{align}
&
\W_{(IJ)KL}(\br,\bs)
\notag\\
=\,&
4\pi
\int_0^{+\infty}dk\,
\int_0^{+\infty}dl\,
\int_{-1}^1d\Theta\,
\int_{-1}^1d\Theta'\,
\int_{-\pi/2}^{\pi/2}d\phi\,k^2\,l^2\,
\notag\\[4pt]&\hskip10mm\times
\bigg\{
 \AOne(k,l,|\bk+\bl|)\, \delta_{IJ}
 \bigg( \delta_{KL}
       +\frac{1}{k^2}\,\frac{\partial^2}{\partial r_K\partial r_L}
       +\frac{1}{l^2}\,\frac{\partial^2}{\partial s_K\partial s_L}
       -\frac{\bk\cdot\bl}{k^2\,l^2}\,\frac{\partial^2}{\partial r_K\partial s_L}
\bigg)
\notag\\[4pt]&\hskip18mm
+\ATwo(k,l,|\bk+\bl|)\,
 \bigg( \delta_{IK}+\frac{1}{k^2}\,\frac{\partial^2}{\partial r_I\partial r_K}\bigg)
 \bigg( \delta_{JL}+\frac{1}{l^2}\,\frac{\partial^2}{\partial s_J\partial s_L}\bigg)
\notag\\[4pt]&\hskip18mm
+\ATwo(k,l,|\bk+\bl|)\,
 \bigg(\delta_{JK}+\frac{1}{k^2}\,\frac{\partial^2}{\partial r_J\partial r_K}\bigg)
 \bigg(\delta_{IL}+\frac{1}{l^2}\,\frac{\partial^2}{\partial s_I\partial s_L}\bigg)
\notag\\[4pt]&\hskip18mm
-\AThree(k,l,|\bk+\bl|)\,\delta_{IJ}
  \bigg(\frac{\partial}{\partial s_K}-\frac{\bk\cdot\bl}{k^2}\,\frac{\partial}{\partial r_K}\bigg)
  \bigg(\frac{\partial}{\partial r_L}-\frac{\bk\cdot\bl}{l^2}\,\frac{\partial}{\partial s_L}\bigg)
\notag\\[4pt]&\hskip18mm
+\AFour(k,l,|\bk+\bl|)
 \bigg(\frac{\partial}{\partial r_I}+\frac{\partial}{\partial s_I}\bigg)
 \bigg(\frac{\partial}{\partial r_J}+\frac{\partial}{\partial s_J}\bigg)
  \bigg(\frac{\partial}{\partial s_K}-\frac{\bk\cdot\bl}{k^2}
  \frac{\partial}{\partial r_K}\bigg)
\notag\\[4pt]&\hskip30mm\times  
  \bigg(\frac{\partial}{\partial r_L}-\frac{\bk\cdot\bl}{l^2}
  \frac{\partial}{\partial s_L}\bigg)
\bigg\} 
\cos(rk\Theta+sl\Theta')
\label{w(ij)kl_IsotropicExpression_ps}
\end{align}
Here,
\begin{align}
\bk\cdot\bl=k\,l\,\Big(\Theta\,\Theta'-\sqrt{1-\Theta^2}\,
\sqrt{1-{\Theta'}^2}\,\sin\phi\Big),\quad
 |\bk+\bl|^2=k^2+l^2+2\,\bk\cdot\bl
\end{align}
are used in \eqref{qwij_IsotropicExpression_ps} through 
\eqref{w(ij)kl_IsotropicExpression_ps}.

The integrals can be evaluated numerically with the help of software packages
 like C{\footnotesize UBA} (\cite{Hahn2005}, \cite{Hahn2006}, \cite{httpcuba}).

\subsection{Objective Function}
\ \ \ \
Within the supposed incompressibility, 
 homogeneity and isotropy, we have set up a mathematical structure with the 
correlations up to the fourth order, without additional approximations involved.
This structure contains three fundamental elements:
 One is the primary dynamical equations of evolution for
$\tW_{kk}$, $\GOne$ and $\GFour$,  \eqref{twkkEvolution} through \eqref{G4Evolution}.
The second is a set of linear equality constraints and a set of
 linear and quadratic inequality constraints listed
 in Subsection \ref{subsec:Constraints}.
The third is the scalar functions $\AOne$, $\ATwo$, $\AThree$ and $\AFour$,
yet to be determined.

Due to the linear forms of the dynamical equations \eqref{twkkEvolution}
through \eqref{G4Evolution},
 $\tW_{kk}(t)$, $\GOne(t)$ and $\GFour(t)$ can be formally solved 
 in terms of $\ATwo(\tau)$, $\tau\leq t$,
under appropriate initial conditions.
As a  consequence, the above-mentioned three elements effectively result 
in a mathematical structure of
 linear and quadratic constraints intrinsic for $\AOne$, 
 $\ATwo$, $\AThree$ and $\AFour$ 
in  incompressible homogeneous isotropic turbulence.

It is known from the closure problem of turbulence that
 this mathematical structure of intrinsic constraints allows 
 many feasible solutions for $\Ai$ to exist,
 under an initial condition given;
and a specific closure tends to reduce the number of solutions.
A conventional closure scheme, like the quasi-normal, 
essentially adds another set of equality constraints,
 and these added ones may not be compatible with the intrinsic,
 as specifically demonstrated in
\cite{OBrienFrancis:1962} and \cite{Ogura:1963}.
Considering the big number and various origins of the  intrinsic constraints,
we expect that any such closures will have the  difficulty to satisfy them.
This observation leads to the option to bypass the conventional:
 one objective function is introduced which is to be optimized 
 under the intrinsic constraints,
 with $\AOne$, $\ATwo$, $\AThree$ and $\AFour$ as the control variables.
That is, the turbulence modeling problem is transformed into an optimal control problem.

The question now is what objective function is to be adopted.
We select tentatively the turbulent energy $\W_{kk}(\bo)$, based on 
the following considerations.
The first is that it reflects the accumulative impact of the third and fourth order 
correlations,
 as indicated by the dynamical equations \eqref{twkkEvolution} through
 \eqref{G4Evolution}.
The second is that it is an invariant trace of the second order correlation tensor
or matrix
 and it may be used to quantify the spread-out of the probability density function 
 of the fluctuations,
 as argued in PART I \cite{TaoRamakrishna2010Part1}.
The third is that it results in a linear (and concave) function of $\AOne$,
 $\ATwo$, $\AThree$ and $\AFour$ to be maximized,
 which have the advantage to be more easily dealt with from a computational perspective,
 unlike any other invariants of the second order correlations.
It then follows that
\begin{align}
\W_{kk}(\bo)
=\int_{\mathbb{R}^3} d\bk\, \tW_{kk}(\bk)
=4\pi\int_{0}^{+\infty} dk\, k^2\,\tW_{kk}(k)
\ \ \text{to be maximized}
 \label{ObjectiveFunction}
\end{align}
 for the problem of our concern here. 
The objective is a linear function of the control variables $\AOne$, $\ATwo$, 
$\AThree$ and $\AFour$,
 when discretized, 
with $\tW_{kk}$, $\GOne$ and $\GFour$ as the primary state variables
 and under the intrinsic constraints listed in Subsection \ref{subsec:Constraints}.
The present optimization problem is a second-order cone programming problem 
\cite{Loboetal1998}.

We mention that there is a thermodynamic setting underlying
the Navier-Stokes equations \eqref{HIT_NSEqsInPhysicalSpace}
in that $\nu\geq 0$ for an incompressible Newtonian fluid
is imposed on the basis of the second law of thermodynamics.
Dealing with the statistical average of the hydrodynamic turbulent fluctuations, 
we construct the present formulation
 from the perspective of information theory regarding the objective function;
The intrinsic constraints result from the mathematical average operation
-- the supposed isotropy and the applications of
the Cauchy-Schwarz inequality and the non-negativity of
the variance of products to the elementary 
correlation functions and the structure functions;
The impact of the Navier-Stokes equations is reflected by the dynamical equations.
We have not included any constraint built on the basis of specific physical models,
considering its potential incompatibility with the intrinsic ones.

\subsection{General Properties}\label{subsec:GeneralProperties}
\ \ \ \
Some mathematical properties of the SOCP problem can be established straight-forwardly.
We have the known relation of
\begin{align}
\frac{\partial}{\partial t}\W_{jj}(\bo)
=2 \,\nu\, \frac{\partial^2}{\partial r_k\partial  r_k}\W_{jj}(\br)\bigg|_{\br=\bo}
=-2 \,\nu\int_{\mathbb{R}^3} d\bk\, k^2\, \tW_{kk}(k)<0
\label{EvolutionOfwkk(r=0)}
\end{align}
following from \eqref{HIT_CLMInPhysicalSpace_2p_r}, \eqref{FourierTransform}, 
\eqref{tQj_Isotropic},
 \eqref{tWijk_Isotropic} and \eqref{twkk_NonNegativeConstraint}.
This monotonic decay is expected physically due to the lacking of external energy
supply and the effect of viscous dissipation.

Next, we have that
\begin{align}
&
 \text{if $\{\AOne,\, \ATwo,\, \AThree,\, \AFour,\, \GOne,\, \GFour,\, \tW_{kk}\}$ 
 is an optimal solution}, 
\notag\\[4pt]&
\text{then $\lambda\,\{\AOne,\, \ATwo,\, \AThree,\, \AFour,\, \GOne,\, \GFour,\,
\tW_{kk}\}$, $\lambda\in(0,1]$,
           is an optimal solution too.}
\label{SOCP_Scaling}
\end{align}
Here, the restriction of the scaling factor $\lambda\leq 1$ comes from
\eqref{VarianceInequality}.

Thirdly, a known scaling property of the Navier-Stokes equations implies
that under the transformation of
\begin{align}
&
 \l\{t,\,\bk,\, \br,\, \AOne,\, \ATwo,\, \AThree,\, \AFour,\, \GOne,\, \GFour,\,
 \tW_{kk}\r\}
\notag\\[4pt]&\hskip10mm
\rightarrow
 \l\{t/\nu,\,\bk,\, \br,\, \nu^4\AOne,\, \nu^4\ATwo,\, \nu^4\AThree,\, \nu^4\AFour,\, 
    \nu^3\GOne,\, \nu^3\GFour,\, \nu^2\tW_{kk}\r\} 
\label{SOCP_nuScaling}
\end{align}
the kinematic viscosity $\nu$ will be removed from the resultant SOCP problem 
whose mathematical equations have the same forms as the original under $\nu=1$.

Compared with the formulation of \cite{ProudmanReid1954} and \cite{Tatsumi1957},
the present is much more complex mathematically and computationally,
 resulting from the different closure strategies underlying.

\section{Asymptotic State Solutions}\label{sec:AsymptoticStateSolutions}
\ \ \ \
To test the SOCP problem above against the experimental and DNS data,
we focus on the asymptotic states of decaying at large time, 
which are characterized within the present framework by certain
conditions to be specified below.

First, as pointed out in \cite{ProudmanReid1954},
 dynamical equations \eqref{G1Evolution} and \eqref{SummationG4Evolution}
can be solved directly for $\GOne(|\bk+\bl|,k,l,t)$ and 
$\GFour(l,|\bk+\bl|,k,t)+\GFour(|\bk+\bl|,k,l,t)+\GFour(k,l,|\bk+\bl|,t)$ 
which decay exponentially with time
 under $\bk\not=\mathbf{0}$ or $\bl\not=\mathbf{0}$.
Consequently, we take these quantities as essentially trivial at large time
to characterize the asymptotic states,
\begin{align}
  \GOne(m,k,l,t)=0,\quad \bm+\bk+\bl=\bo 
\label{G1Evolution_Trivial}
\end{align}
and
\begin{align}
 \GFour(l,m,k,t)+\GFour(m,k,l,t)+\GFour(k,l,m,t)=0,\quad \bl+\bm+\bk=\bo 
\label{SummationG4Evolution_Trivial}
\end{align}
Here, we have removed the singularity of $k=0$ and $l=0$, motivated by
the supposed continuous distributions of $\GOne$ and $\GFour$ in the wave number space
for all time, especially in the limit of $t\rightarrow+\infty$.
Of course, the above two hold if we take $\GOne=0$ and $\GFour=0$ 
as the initial conditions for the artificial transient phase before the asymptotic states
emerge.

Next, from a numerical simulation perspective 
 and the consideration of simplicity,
there is the necessity for the adoption of a finite support for 
 $\AOne$, $\ATwo$, $\AThree$ and $\AFour$, and consequently finite supports for
 $\GFour$ and $\tW_{kk}$ 
following from \eqref{twkkEvolution} and \eqref{G4Evolution};
Let $\maxk$ denote the adopted constant upper bound for $k$, 
which makes it convenient and less time-consuming
the computation of the integrals involved in those constraints
formed in the physical space and the objective.
To normalize the finite supports with $\maxk$ and to remove $\nu$
from the asymptotic state solutions,
we introduce the non-dimensional scalings through
\begin{align}
&
 t=\frac{\hti}{\nu\,(\maxk)^2},\ \ 
\bk=\maxk\,\hbk,\ \ 
\tW_{kk}(k,t)=\frac{\nu^2\,\tWAsy_{kk}(\hk,\hti)}{\maxk},\ \
\GFour(l,m,k,t)=\frac{\nu^3\,\GFourAsy(\hl,\hm,\hk,\hti)}{(\maxk)^4},\ \ 
\notag\\[4pt]&
 \AOne(m,k,l,t)=\frac{\nu^4\,\AOneAsy(\hm,\hk,\hl,\hti)}{(\maxk)^2},\ \
 \ATwo(m,k,l,t)=\frac{\nu^4\,\ATwoAsy(\hm,\hk,\hl,\hti)}{(\maxk)^2},\ \
\notag\\[4pt]&
 \AThree(m,k,l,t)=\frac{\nu^4\,\AThreeAsy(\hm,\hk,\hl,\hti)}{(\maxk)^4},\ \
 \AFour(m,k,l,t)=\frac{\nu^4\,\AFourAsy(\hm,\hk,\hl,\hti)}{(\maxk)^6}
\label{AsymptoticScalingUpdatedInWaveSpace}
\end{align}
which is an extension of \eqref{SOCP_nuScaling}.
It then follows that the correlations in the physical space are scaled through
\begin{align}
&
\br=\frac{\hbr}{\maxk},\ \
\W_{ij}(\br,t)=(\nu\,\maxk)^2\,\WAsy_{ij}(\hbr,\hti),\ \
\overline{\w_{i}\vort_{j}}(\br,t)
=\nu^2(\maxk)^3\,\overline{\w_{i}\vort_{j}}^{(a)}(\hbr,\hti),
\notag\\[4pt]&
\overline{\vort_{i}\vort_{j}}(\br,t)
=\nu^2(\maxk)^4\,\overline{\vort_{i}\vort_{j}}^{(a)}(\hbr,\hti),\ \
\W_{ijk}(\br,\bs,t)=(\nu\,\maxk)^3\,\WAsy_{ijk}(\hbr,\hbs,\hti),
\notag\\[4pt]&
\W_{(IJ)KL}(\br,\bs,t)=(\nu\,\maxk)^4\,\WAsy_{(IJ)KL}(\hbr,\hbs,\hti),\ \
\Q(\br,t)=(\nu\,\maxk)^4\,\QAsy(\hbr,\hti),
\notag\\[4pt]&
\Q_{KL}(\br,\bs,t)=(\nu\,\maxk)^4\,\QAsy_{KL}(\hbr,\hbs,\hti),\ \
\overline{\w_{i}(\bx)\w_{j}(\bx)\vort_{k}(\by)\vort_{l}(\bz)}
=\nu^4(\maxk)^6\,
  \overline{\w_{i}\w_{j}\vort_{k}\vort_{l}}^{(a)}(\hbr,\hbs,\hti),
\notag\\[4pt]&
\overline{\vort_{i}(\bx)\vort_{j}(\bx)\vort_{k}(\by)\vort_{l}(\bz)}
=\nu^4(\maxk)^8\,
\overline{\vort_{i}\vort_{j}\vort_{k}\vort_{l}}^{(a)}(\hbr,\hbs,\hti),
  \ \cdots
\label{AsymptoticScalingUpdatedInPhysicalSpace}
\end{align}
Under the above scalings, \eqref{twkkEvolution} and \eqref{G4Evolution} 
are, respectively, transformed into
\begin{align}
&
 \bigg(\frac{\partial}{\partial \hti}+2\,\hk^2\bigg)\tWAsy_{kk}(\hk,\hti)   
\notag\\[4pt]
=\,&
4\pi \int_0^{1}d\hl \int_{|\hl-\hk|}^{\min(1,\hl+\hk)}d|\hbk+\hbl|\,\hl\,|\hbk+\hbl|\,
       \big(1-\Theta^2\big)
\notag\\[4pt]&\hskip12mm\times
\bigg[
-\bigg(
       2
      +\frac{2\,\hl^2\,\Theta^2+\hk\,\hl\,\Theta-\hl^2}{|\hbk+\hbl|^2}
   \bigg)\,\hk\,\GFourAsy(|\hbk+\hbl|,\hl,\hk,\hti)
\notag\\[4pt]&\hskip20mm
+\bigg(
 \Theta
+\frac{\hk\,\hl\,(1-\Theta^2)}{|\hbk+\hbl|^2}
   \bigg)\,\hl\,\GFourAsy(|\hbk+\hbl|,\hk,\hl,\hti)
\bigg],
\ \   
\Theta=\frac{|\hbk+\hbl|^2-\hk^2-\hl^2}{2\,\hk\,\hl}
\label{twkkEvolution_ScalingUpdated}
\end{align}
and
\begin{align}
 \bigg(\frac{\partial }{\partial \hti}+\hk^2+\hl^2+|\hbk+\hbl|^2\bigg)
 \GFourAsy(\hl,|\hbk+\hbl|,\hk,\hti)
=
 \ATwoAsy(|\hbk+\hbl|,\hk,\hl,\hti)-\ATwoAsy(|\hbk+\hbl|,\hl,\hk,\hti)
\label{G4Evolution_ScalingUpdated}
\end{align}
All the constraints listed in Subsection \ref{subsec:Constraints}
and constraint \eqref{SummationG4Evolution_Trivial} maintain the same structures 
after the transformation -- the mere replacement
of the original dimensional quantities with the corresponding dimensionless. 
The definition of support itself implies that
\begin{align}
\AjAsy(\hk,\hm,\hl,\hti)=\GFourAsy(\hk,\hm,\hl,\hti)=0,
\ \ \hk\geq 1\ \text{or}\ \hm\geq 1\ 
\text{or}\ \hl\geq 1;\ \ 
\tWAsy_{kk}(\hk,\hti)=0,\ \hk\geq 1
\label{SupportConstraints_ScalingUpdated} 
\end{align}
Here, the inclusion of $>1$ is to account for the operations
 like $\hbk\rightarrow-\hbk-\hbl$, $\hbm=\hbk+\hbl\rightarrow-\hbk$, etc.
The objective function in \eqref{ObjectiveFunction} takes the equivalent form of
\begin{align}
\int_{0}^{1} d\hk\, \hk^2\,\tWAsy_{kk}(\hk,\hti)
\ \ \text{to be maximized at each and every great $\hti$}
\label{ObjectiveFunction_ScalingUpdated}
\end{align}
We have thus a transformed SOCP problem for the asymptotic states
 with the magnitude of its wave-numbers bounded by $1$ and parameter-free.

Thirdly, under the above scaling,  \eqref{EvolutionOfwkk(r=0)} becomes
\begin{align}
\frac{\partial}{\partial \hti}\WAsy_{jj}(\bo,\hti)
=-8\pi\int_{0}^{1} d\hk\, \hk^4\, \tWAsy_{kk}(\hk,\hti)<0
\label{EvolutionOfwkk(r=0)_ScalingUpdated}
\end{align}
The question is whether the linear forms of
 \eqref{twkkEvolution_ScalingUpdated} and \eqref{G4Evolution_ScalingUpdated}
and the decaying behavior of \eqref{EvolutionOfwkk(r=0)_ScalingUpdated}
result in the automatic satisfaction of
the variance inequalities \eqref{VarianceInequality}
 at large time in the asymptotic states. 

Computationally, we need to integrate \eqref{twkkEvolution_ScalingUpdated} and
\eqref{G4Evolution_ScalingUpdated} starting
from $\hti=0$ with appropriately prescribed initial conditions for $\tWAsy_{kk}$ and $\GFourAsy$
that satisfies the constraints \eqref{twkk_NonNegativeConstraint}
and \eqref{tWijk_Isotropic_Constraints},
and there is an artificial transient phase before the emergence of the asymptotic states.

\subsection{Sub-model}
\ \ \ \
For the numerical simulation of the asymptotic states,  
we need to discretize the unit cube support $\SupportAsy=[0,1]^3$ for 
 $\AOneAsy$, $\ATwoAsy$, $\AThreeAsy$, $\AFourAsy$ and $\GFourAsy$,
 the constraints and  \eqref{twkkEvolution_ScalingUpdated}
 and \eqref{G4Evolution_ScalingUpdated},
and we obtain a large-scale SOCP problem.
To have a rough estimate about the number of discrete variables involved,
 we take a uniform mesh size of $1/50$ along each edge of the cube which results in
 $51^3$ nodes for the mesh, and accordingly, there are $51^3\times 5$ discrete variables
 associated with $\AjAsy$ and $\GFourAsy$ at each discretized time instant.
Meanwhile, there are a large number of the discretized constraints to be imposed.
Based on this estimate and the computing resources available, 
we will focus on the asymptotic states of
 the sub-model which involves only the second and the third order 
correlations,
 \eqref{tWijk_Isotropic_Constraints},
 \eqref{twkk_NonNegativeConstraint},
 \eqref{StructureFunction_wwIneq_SOC} through \eqref{StructureFunction_vvIneq_SOC},
 \eqref{SummationG4Evolution_Trivial},
 \eqref{twkkEvolution_ScalingUpdated},
 \eqref{SupportConstraints_ScalingUpdated}
and \eqref{ObjectiveFunction_ScalingUpdated},
which are recorded below.
\begin{align}
&
 \bigg(\frac{\partial}{\partial \hti}+2\,\hk^2\bigg)\tWAsy_{kk}(\hk,\hti)   
\notag\\[4pt]
=\,&
4\pi \int_0^{1}d\hl \int_{|\hl-\hk|}^{\min(1,\hl+\hk)}d\hm\,\hl\,\hm\,
       \big(1-\Theta^2\big)
\notag\\[4pt]&\hskip12mm\times
\bigg[
-\bigg(
       2
      +\frac{2\,\hl^2\,\Theta^2+\hk\,\hl\,\Theta-\hl^2}{\hm^2}
   \bigg)\,\hk\,\GFourAsy(\hm,\hl,\hk,\hti)
\notag\\[4pt]&\hskip20mm
+\bigg(
 \Theta
+\frac{\hk\,\hl\,(1-\Theta^2)}{\hm^2}
   \bigg)\,\hl\,\GFourAsy(\hm,\hk,\hl,\hti)
\bigg],
\ \   
\Theta=\frac{\hm^2-\hk^2-\hl^2}{2\,\hk\,\hl}
\label{twkkEvolutionRd_Submodel}
\end{align}
\begin{align}
 \tWAsy_{kk}(\hk,\hti)\geq 0
\label{twkk_NonNegative_Submodel}
\end{align}
\begin{align}
&
\Big[
 \WAsy_{ij}(\hbs'-\hbr,\hti)
-\WAsy_{ij}(\hbs-\hbr,\hti)
-\WAsy_{ij}(\hbs',\hti)
+\WAsy_{ij}(\hbs,\hti)
 \notag\\[4pt]&\hskip5mm
+2\,\alpha \Big(\WAsy_{ij}(\bo,\hti)-\WAsy_{ij}(\hbs'-\hbs,\hti)\Big)
-2\,\beta \Big(\WAsy_{ij}(\bo,\hti)-\WAsy_{ij}(\hbr,\hti)\Big)
 \notag\\[4pt]&\hskip5mm
-\alpha\beta \Big(
 \WAsy_{ij}(\hbs'-\hbr,\hti)
-\WAsy_{ij}(\hbs-\hbr,\hti)
-\WAsy_{ij}(\hbs',\hti)
+\WAsy_{ij}(\hbs,\hti)
 \Big)
 \Big]^2
\notag\\[4pt]
\leq\,&
4\,\Big[
 \WAsy_{\underlinei\,\underlinei}(\bo,\hti)
-\WAsy_{\underlinei\,\underlinei}(\hbr,\hti)
+\alpha^2 \Big(
 \WAsy_{\underlinei\,\underlinei}(\bo,\hti)
-\WAsy_{\underlinei\,\underlinei}(\hbs'-\hbs,\hti)
 \Big)
 \notag\\[4pt]&\hskip10mm
+\alpha \Big(
 \WAsy_{\underlinei\,\underlinei}(\hbs'-\hbr,\hti)
-\WAsy_{\underlinei\,\underlinei}(\hbs-\hbr,\hti)
-\WAsy_{\underlinei\,\underlinei}(\hbs',\hti)
+\WAsy_{\underlinei\,\underlinei}(\hbs,\hti)
\Big)
\Big]
\notag\\[4pt]&\times
\Big[
 \WAsy_{\underlinej\,\underlinej}(\bo,\hti)
-\WAsy_{\underlinej\,\underlinej}(\hbs'-\hbs,\hti)
+\beta^2 \Big(
 \WAsy_{\underlinej\,\underlinej}(\bo,\hti)
-\WAsy_{\underlinej\,\underlinej}(\hbr,\hti)
 \Big)
 \notag\\[4pt]&\hskip10mm
-\beta \Big(
 \WAsy_{\underlinej\,\underlinej}(\hbs'-\hbr,\hti)
-\WAsy_{\underlinej\,\underlinej}(\hbs-\hbr,\hti)
-\WAsy_{\underlinej\,\underlinej}(\hbs',\hti)
+\WAsy_{\underlinej\,\underlinej}(\hbs,\hti)
\Big)
\Big]
   ,\ \ ij=11,12
\label{wwIneq_Submodel}
\end{align}
\begin{align}
&
\Big[
 \overline{\w_i\Vorticity_j}^{(a)}(\hbs'-\hbr,\hti)
-\overline{\w_i\Vorticity_j}^{(a)}(\hbs-\hbr,\hti)
-\overline{\w_i\Vorticity_j}^{(a)}(\hbs',\hti)
+\overline{\w_i\Vorticity_j}^{(a)}(\hbs,\hti)
 \notag\\[4pt]&\hskip5mm
+\alpha\beta \Big(
 \overline{\w_i\Vorticity_j}^{(a)}(\hbs'-\hbr,\hti)
-\overline{\w_i\Vorticity_j}^{(a)}(\hbs-\hbr,\hti)
-\overline{\w_i\Vorticity_j}^{(a)}(\hbs',\hti)
+\overline{\w_i\Vorticity_j}^{(a)}(\hbs,\hti)
 \Big)
 \Big]^2
\notag\\[4pt]
\leq\,&
4\,\Big[
 \WAsy_{\underlinei\,\underlinei}(\bo,\hti)
-\WAsy_{\underlinei\,\underlinei}(\hbr,\hti)
+\alpha^2 \Big(
 \WAsy_{\underlinei\,\underlinei}(\bo,\hti)
-\WAsy_{\underlinei\,\underlinei}(\hbs'-\hbs,\hti)
 \Big)
 \notag\\[4pt]&\hskip10mm
+\alpha \Big(
 \WAsy_{\underlinei\,\underlinei}(\hbs'-\hbr,\hti)
-\WAsy_{\underlinei\,\underlinei}(\hbs-\hbr,\hti)
-\WAsy_{\underlinei\,\underlinei}(\hbs',\hti)
+\WAsy_{\underlinei\,\underlinei}(\hbs,\hti)
\Big)
\Big]
\notag\\[4pt]&\times
\Big[
 \overline{\Vorticity_{\underlinej}\Vorticity_{\underlinej}}^{(a)}(\bo,\hti)
-\overline{\Vorticity_{\underlinej}\Vorticity_{\underlinej}}^{(a)}(\hbs'-\hbs,\hti)
+\beta^2 \Big(
 \overline{\Vorticity_{\underlinej}\Vorticity_{\underlinej}}^{(a)}(\bo,\hti)
-\overline{\Vorticity_{\underlinej}\Vorticity_{\underlinej}}^{(a)}(\hbr,\hti)
 \Big)
 \notag\\[4pt]&\hskip10mm
-\beta \Big(
 \overline{\Vorticity_{\underlinej}\Vorticity_{\underlinej}}^{(a)}(\hbs'-\hbr,\hti)
-\overline{\Vorticity_{\underlinej}\Vorticity_{\underlinej}}^{(a)}(\hbs-\hbr,\hti)
-\overline{\Vorticity_{\underlinej}\Vorticity_{\underlinej}}^{(a)}(\hbs',\hti)
+\overline{\Vorticity_{\underlinej}\Vorticity_{\underlinej}}^{(a)}(\hbs,\hti)
\Big)
\Big],\ \ ij=11,12
\label{wvIneq_Submodel}
\end{align}
\begin{align}
&
\Big[
 \overline{\Vorticity_i\Vorticity_j}^{(a)}(\hbs'-\hbr,\hti)
-\overline{\Vorticity_i\Vorticity_j}^{(a)}(\hbs-\hbr,\hti)
-\overline{\Vorticity_i\Vorticity_j}^{(a)}(\hbs',\hti)
+\overline{\Vorticity_i\Vorticity_j}^{(a)}(\hbs,\hti)
 \notag\\[4pt]&\hskip5mm
+2\,\alpha \Big(
 \overline{\Vorticity_i\Vorticity_j}^{(a)}(\bo,\hti)
-\overline{\Vorticity_i\Vorticity_j}^{(a)}(\hbs'-\hbs,\hti)
\Big)
-2\,\beta \Big(
 \overline{\Vorticity_i\Vorticity_j}^{(a)}(\bo,\hti)
-\overline{\Vorticity_i\Vorticity_j}^{(a)}(\hbr,\hti)
 \Big)
 \notag\\[4pt]&\hskip5mm
-\alpha\beta \Big(
 \overline{\Vorticity_i\Vorticity_j}^{(a)}(\hbs'-\hbr,\hti)
-\overline{\Vorticity_i\Vorticity_j}^{(a)}(\hbs-\hbr,\hti)
-\overline{\Vorticity_i\Vorticity_j}^{(a)}(\hbs',\hti)
+\overline{\Vorticity_i\Vorticity_j}^{(a)}(\hbs,\hti)
 \Big)
 \Big]^2
\notag\\[4pt]
\leq\,&
4\,\Big[
 \overline{\Vorticity_{\underlinei}\Vorticity_{\underlinei}}^{(a)}(\bo,\hti)
-\overline{\Vorticity_{\underlinei}\Vorticity_{\underlinei}}^{(a)}(\hbr,\hti)
+\alpha^2 \Big(
 \overline{\Vorticity_{\underlinei}\Vorticity_{\underlinei}}^{(a)}(\bo,\hti)
-\overline{\Vorticity_{\underlinei}\Vorticity_{\underlinei}}^{(a)}(\hbs'-\hbs,\hti)
 \Big)
 \notag\\[4pt]&\hskip10mm
+\alpha \Big(
 \overline{\Vorticity_{\underlinei}\Vorticity_{\underlinei}}^{(a)}(\hbs'-\hbr,\hti)
-\overline{\Vorticity_{\underlinei}\Vorticity_{\underlinei}}^{(a)}(\hbs-\hbr,\hti)
-\overline{\Vorticity_{\underlinei}\Vorticity_{\underlinei}}^{(a)}(\hbs',\hti)
+\overline{\Vorticity_{\underlinei}\Vorticity_{\underlinei}}^{(a)}(\hbs,\hti)
\Big)
\Big]
\notag\\[4pt]&\times
\Big[
 \overline{\Vorticity_{\underlinej}\Vorticity_{\underlinej}}^{(a)}(\bo,\hti)
-\overline{\Vorticity_{\underlinej}\Vorticity_{\underlinej}}^{(a)}(\hbs'-\hbs,\hti)
+\beta^2 \Big(
 \overline{\Vorticity_{\underlinej}\Vorticity_{\underlinej}}^{(a)}(\bo,\hti)
-\overline{\Vorticity_{\underlinej}\Vorticity_{\underlinej}}^{(a)}(\hbr,\hti)
 \Big)
 \notag\\[4pt]&\hskip10mm
-\beta \Big(
 \overline{\Vorticity_{\underlinej}\Vorticity_{\underlinej}}^{(a)}(\hbs'-\hbr,\hti)
-\overline{\Vorticity_{\underlinej}\Vorticity_{\underlinej}}^{(a)}(\hbs-\hbr,\hti)
-\overline{\Vorticity_{\underlinej}\Vorticity_{\underlinej}}^{(a)}(\hbs',\hti)
+\overline{\Vorticity_{\underlinej}\Vorticity_{\underlinej}}^{(a)}(\hbs,\hti)
\Big)
\Big],\ \  ij=11,12
\label{vvIneq_Submodel}
\end{align}
\begin{align}
\GFourAsy(\hm,\hk,\hl,\hti)+\GFourAsy(\hl,\hk,\hm,\hti)=0,\quad \hbm+\hbk+\hbl=\bo 
\label{G4Symmetry_Submodel}
\end{align}
\begin{align}
 \GFourAsy(\hl,\hm,\hk,\hti)
+\GFourAsy(\hm,\hk,\hl,\hti)
+\GFourAsy(\hk,\hl,\hm,\hti)=0,\quad \hbl+\hbm+\hbk=\bo
\label{SummationG4Evolution_Submodel}
\end{align}
\begin{align}
\GFourAsy(\hk,\hm,\hl,\hti)=0,\ \ \hk\geq 1\ \text{or}\ \hm\geq 1\ \text{or}\ \hl\geq 1
\label{SupportConstraints_Submodel} 
\end{align}
\begin{align}
\l|\GFourAsy(\hk,\hm,\hl,\hti)\r|\leq 1
\label{G2Bound_Submodel}
\end{align}
and
\begin{align}
\int_{0}^{1} d\hk\, \hk^2\,\tWAsy_{kk}(\hk,\hti)
\ \ \text{to be maximized at each and every great $\hti$}
\label{ObjectiveFunction_Submodel}
\end{align}
Here, the addition of \eqref{G2Bound_Submodel} is to preserve
the scaling property of \eqref{SOCP_Scaling} for the sub-model,
due to the non-enforceability of \eqref{VarianceInequality};
The upper bound is normalized to the unity,
allowed by the scaling property of the equations involved;
This bound restricts the feasible domain, and thus, it plays the role of reducing computing time.
The above is a SOCP problem when discretized.
We should mention that within the sub-model,
 the equality constraint \eqref{tQj_Isotropic} is satisfied automatically
which lends the basis to take \eqref{twkkEvolutionRd_Submodel}.
Also, the general property, \eqref{EvolutionOfwkk(r=0)}
or \eqref{EvolutionOfwkk(r=0)_ScalingUpdated}, holds within the sub-model.

The constraints of \eqref{G4Symmetry_Submodel} 
and \eqref{SummationG4Evolution_Submodel} can be simplified for the sake
of numerical simulation.
Under \eqref{G4Symmetry_Submodel}, the equality constraint 
 \eqref{SummationG4Evolution_Submodel} reduces to and is equivalent to
\begin{align}
 \GFourAsy(\hl,\hm,\hk,\hti)
+\GFourAsy(\hm,\hk,\hl,\hti)
+\GFourAsy(\hk,\hl,\hm,\hti)=0,\ \
\hk<\hl<\hm\leq \hk+\hl,\ \
\hbm=-\hbk-\hbl
\label{SummationG4Evolution_SubmodelSubcase03}
\end{align}
To prove this claim,
 we first infer from \eqref{G4Symmetry_Submodel} the following two specific 
 representations
 compatible with \eqref{SupportConstraints_Submodel},
\begin{align}
\GFourAsy(\hm,\hk,\hl,\hti)+\GFourAsy(\hl,\hk,\hm,\hti)=0,\ \ 
\big|\hk-\hl\big|\leq\hm\leq\min\s\big\{\hk+\hl,1\big\},\ \
0\leq\hk, \hl\leq 1
\label{G4Symmetry_Submodel_b}
\end{align}
and
\begin{align}
&
\GFourAsy(\hl,\hk,\hl,\hti)=0,\ \ 0\leq\hk\leq\min\s\big\{2\,\hl,1\big\},
\ \ 0\leq\hl\leq 1;
\notag\\[4pt]&
\GFourAsy(\hm,\hk,\hl,\hti)+\GFourAsy(\hl,\hk,\hm,\hti)=0,\ \ 
\hm-\hl\leq\hk\leq\min\s\big\{\hm+\hl,1\big\},\ \
0\leq\hl<\hm\leq 1
\label{G4Symmetry_Submodel_aa}
\end{align}
using $|\hk-\hl|\leq|\hbk+\hbl|\leq\hk+\hl$ and the like.
Next, we verify the equivalence between \eqref{SummationG4Evolution_Submodel}
and \eqref{SummationG4Evolution_SubmodelSubcase03}
under \eqref{G4Symmetry_Submodel},
   \eqref{G4Symmetry_Submodel_b} and \eqref{G4Symmetry_Submodel_aa}
in the three cases  of $\hm=\hl$, $\hm>\hl$ and $\hm<\hl$, respectively.
The case, $\hm=\hl$, is trivial in that 
\eqref{SummationG4Evolution_Submodel} is satisfied
 by \eqref{G4Symmetry_Submodel_b} and (\ref{G4Symmetry_Submodel_aa}$)_1$ if $\hk\geq\hl$
and by \eqref{G4Symmetry_Submodel_aa} if $\hk<\hl$.
The case, $\hm>\hl$, includes three subcases: $\hk=\hl$, $\hk<\hl$ and $\hk>\hl$.
For $\hk=\hl$, \eqref{SummationG4Evolution_Submodel} is satisfied
by \eqref{G4Symmetry_Submodel_aa}. 
For $\hk<\hl$, \eqref{SummationG4Evolution_Submodel} is satisfied by
\eqref{SummationG4Evolution_SubmodelSubcase03}.
Subcase $\hk>\hl$ contains two possibilities of $\hm\geq\hk>\hl$ and $\hk\geq\hm>\hl$:
the first possibility is equivalent to Subcase $\hk<\hl$, 
    which can be demonstrated with \eqref{G4Symmetry_Submodel} and 
    \eqref{G4Symmetry_Submodel_aa}; 
the second possibility is equivalent to the first. 
The last case, $\hm<\hl$, is equivalent to the case of $\hm>\hl$, which can be 
verified with the application of
 \eqref{G4Symmetry_Submodel} to \eqref{SummationG4Evolution_Submodel}.

We can argue for the equivalence
 between \eqref{G4Symmetry_Submodel_b} and \eqref{G4Symmetry_Submodel_aa} based on that 
 \eqref{G4Symmetry_Submodel_aa} can be rewritten in the equivalent form of
\begin{align}
\GFourAsy(\hm,\hk,\hl,\hti)+\GFourAsy(\hl,\hk,\hm,\hti)=0,\ \ 
\hm-\hl\leq\hk\leq\min\s\big\{\hm+\hl,1\big\},\ \
0\leq\hl,\hm\leq 1
\label{G4Symmetry_Submodel_a}
\end{align}
The equivalence between \eqref{G4Symmetry_Submodel} and \eqref{G4Symmetry_Submodel_b} 
comes from that the latter is a specific representation of the former 
and the latter implies the former in that
for any $\hbm$, $\hbk$ and $\hbl$ with $\hbm+\hbk+\hbl=\bo$ involved in
\eqref{G4Symmetry_Submodel},
  $\hbm=-(\hbk+\hbl)$ and $\hm=|\hbk+\hbl|\in[|\hk-\hl|,\hk+\hl]$,
 there exist a corresponding set of $\{\hk,\hl,\hm\}$ in \eqref{G4Symmetry_Submodel_b}.
 Consequently,  \eqref{G4Symmetry_Submodel} and \eqref{G4Symmetry_Submodel_aa} are equivalent.

The above results imply that
 we can replace \eqref{G4Symmetry_Submodel} and \eqref{SummationG4Evolution_Submodel}
with \eqref{SummationG4Evolution_SubmodelSubcase03} and \eqref{G4Symmetry_Submodel_aa}
in numerical simulations.

At every fixed time instant $\hti$,
each of the constraints, \eqref{wwIneq_Submodel} through \eqref{vvIneq_Submodel},
involves a scalar function of the components of $\hbr$, $\hbs$,
$\hbs'$ and their differences and the non-negative constants $\alpha$ and $\beta$
 to be fixed.
To make the SOCP problem computationally feasible,
under specifically fixed $\alpha$ and $\beta$,
an adequate set of the collocation points in the physical space has to
be selected at which the constraints are to be imposed.

\subsection{Discretization Scheme}
\ \ \ \
To carry out the numerical simulation,
we discretize the unit cubic support $[0,1]^3$ with a structured hexahedral 
mesh of uniform size $\deltak$,
\begin{align}
{\cal D}_{\GFourAsy}
=\,&
\bigcup_{n_1=1}^{N-1} 
\bigcup_{n_2=1}^{N-1} 
\bigcup_{n_3=1}^{N-1} 
{\cal H}(n_1,n_2,n_3),
\notag\\[4pt]&
{\cal H}(n_1,n_2,n_3)=
[k_{n_1},k_{n_1}+\deltak)\times
      [k_{n_2},k_{n_2}+\deltak)\times
      [k_{n_3},k_{n_3}+\deltak)
\end{align}
The half-open intervals are adopted to avoid double counting 
in the evaluation of $\GFourAsy$;
 the end interval of each axis should be closed to represent fully the closed support, 
which is not crucial in practice due to the enforcement of
\eqref{SupportConstraints_Submodel}.

The distribution of $\GFourAsy(\hm,\hk,\hl,\hti)$ in cubic cell ${\cal H}(n_1,n_2,n_3)$ 
 is approximated by the tri-linear distribution,
\begin{align}
&
 \GTriLinear(\hm,\hk,\hl,\hti;n_1,n_2,n_3)
\notag\\[4pt]
=
\,&
 \GNodes(n_1+1,n_2+1,n_3+1,\hti)\,
 \frac{\hm-k_{n_1}}{\deltak}\,
 \frac{\hk-k_{n_2}}{\deltak}\,
 \frac{\hl-k_{n_3}}{\deltak}
\notag\\[4pt]&
+
 \GNodes(n_1+1,n_2,n_3+1,\hti)\,
 \frac{\hm-k_{n_1}}{\deltak}\,
 \frac{k_{n_2+1}-\hk}{\deltak}\,
 \frac{\hl-k_{n_3}}{\deltak}
\notag\\[4pt]&
+
 \GNodes(n_1,n_2+1,n_3+1,\hti)\,
 \frac{k_{n_1+1}-\hm}{\deltak}\,
 \frac{\hk-k_{n_2}}{\deltak}\,
 \frac{\hl-k_{n_3}}{\deltak}
\notag\\[4pt]&
+
 \GNodes(n_1,n_2,n_3+1,\hti)\,
 \frac{k_{n_1+1}-\hm}{\deltak}\,
 \frac{k_{n_2+1}-\hk}{\deltak}\,
 \frac{\hl-k_{n_3}}{\deltak}
\notag\\[4pt]&
+
 \GNodes(n_1+1,n_2+1,n_3,\hti)\,
 \frac{\hm-k_{n_1}}{\deltak}\,
 \frac{\hk-k_{n_2}}{\deltak}\,
 \frac{k_{n_3+1}-\hl}{\deltak}
\notag\\[4pt]&
+
 \GNodes(n_1+1,n_2,n_3,\hti)\,
 \frac{\hm-k_{n_1}}{\deltak}\,
 \frac{k_{n_2+1}-\hk}{\deltak}\,
 \frac{k_{n_3+1}-\hl}{\deltak}
\notag\\[4pt]&
+
 \GNodes(n_1,n_2+1,n_3,\hti)\,
 \frac{k_{n_1+1}-\hm}{\deltak}\,
 \frac{\hk-k_{n_2}}{\deltak}\,
 \frac{k_{n_3+1}-\hl}{\deltak}
\notag\\[4pt]&
+
 \GNodes(n_1,n_2,n_3,\hti)\,
 \frac{k_{n_1+1}-\hm}{\deltak}\,
 \frac{k_{n_2+1}-\hk}{\deltak}\,
 \frac{k_{n_3+1}-\hl}{\deltak}
\label{TriLinearApproximationInCell}
\end{align}
where $\GNodes(n_1,n_2,n_3,\hti)$ denotes the value of $\GFourAsy$ 
at node $(n_1,n_2,n_3)$
 at time instant $\hti$.
 The distribution of $\GFourAsy(\hm,\hk,\hl,\hti)$ in ${\cal D}_{\GFourAsy}$
 is approximated by
\begin{align}
\GFourAsy(\hm,\hk,\hl,\hti)
=
\sum_{n_1,\,n_2,\,n_3\,=\,1}^{N-1}
\chi_{[k_{n_1},\, k_{n_1+1})}(\hm)\,
\chi_{[k_{n_2},\, k_{n_2+1})}(\hk)\,
\chi_{[k_{n_3},\, k_{n_3+1})}(\hl)\,
      \GTriLinear(\hm,\hk,\hl,\hti;n_1,n_2,n_3)
\label{TriLinearApproximation}
\end{align}
where $\chi_{[k_{n_1},\, k_{n_1+1})}$ and the like are the characteristic functions.

The direct constraints for $\GNodes(n_1,n_2,n_3,\hti)$ can be established by
the substitution of \eqref{TriLinearApproximationInCell} into
\eqref{SupportConstraints_Submodel},
\eqref{G2Bound_Submodel},
   \eqref{SummationG4Evolution_SubmodelSubcase03} and \eqref{G4Symmetry_Submodel_aa}.
First, the boundary conditions of the finite support \eqref{SupportConstraints_Submodel}
require that
\begin{align}
\GNodes(n_1,n_2,n_3,\hti)=0,\ \ n_1=N\ \text{or}\ n_2=N\ \text{or}\ n_3=N
\label{G4_Symmetry/FiniteSupport}
\end{align}
Next, we evaluate the specific representation of symmetry
 \eqref{G4Symmetry_Submodel_aa} at the nodes and the mid-points
 between the nodes to obtain
\begin{align}
&
 \GNodes(n_1,n_2,n_1,\hti)=0,\ \  1 \leq n_2 \leq \min\{2\,n_1, N-1\},
 \ 1\leq  n_1\leq N-1;
\notag\\[4pt]&
 \GNodes(n_1,n_2,n_3,\hti)+\GNodes(n_3,n_2,n_1,\hti)=0,
\notag\\[4pt]&
  n_3-n_1 \leq n_2 \leq \min\{n_1+n_3, N-1\},\ 1\leq  n_1< n_3\leq N-1
\label{G4_SymmetryIa4}
\end{align}
Thirdly, the evaluations of the asymptotic state constraint
\eqref{SummationG4Evolution_SubmodelSubcase03} 
 at the nodes and the mid-points between the nodes result in
\begin{align}
&
 \GNodes(n_1,n_2,n_3,\hti)
+\GNodes(n_2,n_3,n_1,\hti)
+\GNodes(n_3,n_1,n_2,\hti)
=0,
\notag\\[4pt]&
1\leq n_1 \leq n_2+1,\ \ n_2 \leq n_3+1,\ \ n_3\leq \min\{n_1+n_2,N-1\}
\label{SummationG4Evolution_Discretized_E}
\end{align}
Finally, the bound constraint \eqref{G2Bound_Submodel} leads to
\begin{align}
\l|\GNodes(n_1,n_2,n_3,\hti)\r|\leq 1
\label{G2Bound_Discretized}
\end{align}

To implement constraints \eqref{twkk_NonNegative_Submodel}
  through \eqref{vvIneq_Submodel},
we need to represent $\tWAsy_{kk}$ in terms of $\GFourAsy$
by solving \eqref{twkkEvolutionRd_Submodel}.
To this end, we integrate \eqref{twkkEvolutionRd_Submodel} in the time interval
 $\tau\in[\hti,\hti+\delta\hti]$ to get 
\begin{align}
&
 \tWAsy_{kk}(\hk,\hti+\delta\hti)
\notag\\[4pt]
=\,&
 \exp\s\big(-2\,\hk^2\,\delta\hti\big)\,\tWAsy_{kk}(\hk,\hti)
\notag\\[4pt]&
+4\pi\s
 \int_0^{1}d\hl \int_{|\hl-\hk|}^{\min(1,\hl+\hk)}d\hm\,\hl\,\hm\,
       \big(1-\Theta^2\big)
\notag\\[4pt]&\hskip12mm\times
\bigg[
-\bigg(
       2
      +\frac{2\,\hl^2\,\Theta^2+\hk\,\hl\,\Theta-\hl^2}{\hm^2}
   \bigg)\,\hk
      \int_{\hti}^{\hti+\delta\hti}d\tau\,\exp\s\big[2\,\hk^2
      \big(\tau-\hti-\delta\hti\big)\big]
              \GFourAsy(\hm,\hl,\hk,\tau)
\notag\\[4pt]&\hskip20mm
+\bigg(
 \Theta
+\frac{\hk\,\hl\,(1-\Theta^2)}{\hm^2}
   \bigg)\,\hl
     \int_{\hti}^{\hti+\delta\hti}d\tau\,
            \exp\s\big[2\,\hk^2 \big(\tau-\hti-\delta\hti\big)\big]
              \GFourAsy(\hm,\hk,\hl,\tau)
\bigg],
\notag\\[4pt]&
\Theta=\frac{\hm^2-\hk^2-\hl^2}{2\,\hk\,\hl}
\label{twkkEvolutionRd_Integrated t2tdeltat}
\end{align}
we then approximate $\GFourAsy(\hm,\hk,\hl,\tau)$ in a linear fashion,
\begin{align}
 \GFourAsy(\hm,\hk,\hl,\tau)
=\,&
 \GFourAsy(\hm,\hk,\hl,\hti)\,\frac{\hti+\delta\hti-\tau}{\delta\hti}
+\GFourAsy(\hm,\hk,\hl,\hti+\delta\hti)\,\frac{\tau-\hti}{\delta\hti},\ \
\tau\in\big[\hti,\hti+\delta\hti\big]
\end{align}
and integrate $\int_{\hti}^{\hti+\delta\hti} d\tau$ analytically to obtain
\begin{align}
&
 \tWAsy_{kk}(\hk,\hti+\delta\hti)
\notag\\[4pt]
=\,&
 \exp(-2\,\hk^2\,\delta\hti)\,\tWAsy_{kk}(\hk,\hti)
\notag\\[4pt]&
-\frac{\pi}{16\,\delta\hti}\,\frac{1-(1+2\,\hk^2\,\delta\hti)\,
\exp(-2\,\hk^2\,\delta\hti)}{\hk^7}\,
\notag\\[4pt]&\hskip10mm\times
 \int_0^{1}\frac{d\hl}{\hl} \int_{|\hl-\hk|}^{\min(1,\hl+\hk)}\frac{d\hm}{\hm}\,
                   \Big(
 \hm^4
+\hk^4
+\hl^4
-2\,\hk^2\,\hl^2
-2\,\hk^2\,\hm^2
-2\,\hl^2\,\hm^2
\Big)\,
\notag\\[4pt]&\hskip32mm\times
\bigg[
 \Big(
 \hm^4
-\hk^4
-\hl^4
+2\,\hk^2\,\hl^2
\Big)\,
    \GFourAsy(\hm,\hk,\hl,\hti)
\notag\\[4pt]&\hskip40mm
-\Big(
 \hm^4
+\hl^4
-\hk^2\,\hl^2
+3\,\hk^2\,\hm^2
-2\,\hl^2\,\hm^2
\Big)\,
   2\,\GFourAsy(\hm,\hl,\hk,\hti)
\bigg]
\notag\\[4pt]&
+\frac{\pi}{16\,\delta\hti}\,\frac{1-2\,\hk^2\,\delta\hti
-\exp(-2\,\hk^2\,\delta\hti)}{\hk^7}\,
\notag\\[4pt]&\hskip10mm\times
 \int_0^{1}\frac{d\hl}{\hl} \int_{|\hl-\hk|}^{\min(1,\hl+\hk)}\frac{d\hm}{\hm}\,
                   \Big(
 \hm^4
+\hk^4
+\hl^4
-2\,\hk^2\,\hl^2
-2\,\hk^2\,\hm^2
-2\,\hl^2\,\hm^2
\Big)\,
\notag\\[4pt]&\hskip32mm\times
\bigg[
 \Big(
 \hm^4
-\hk^4
-\hl^4
+2\,\hk^2\,\hl^2
\Big)\,
    \GFourAsy(\hm,\hk,\hl,\hti+\delta\hti)
\notag\\[4pt]&\hskip40mm
-\Big( 
 \hm^4
+\hl^4
-\hk^2\,\hl^2
+3\,\hk^2\,\hm^2
-2\,\hl^2\,\hm^2
\Big)\,
   2\,\GFourAsy(\hm,\hl,\hk,\hti+\delta\hti)
\bigg]
\label{twkkEvolutionRd_tAndtPlus}
\end{align}
Next, substituting \eqref{TriLinearApproximation} into \eqref{twkkEvolutionRd_tAndtPlus}
and integrating
$\int_{\max\{|\hl-\hk|, \hk_{n_1}\}}^{\min\{1, \hl+\hk, \hk_{n_1+1}\}} d\hm$ 
analytically (if desired) result in
\begin{align}
&
 \tWAsy_{kk}(\hk,\hti+\delta\hti)
\notag\\[4pt]
=\,&
 \exp(-2\,\hk^2\,\delta\hti)\,\tWAsy_{kk}(\hk,\hti)
\notag\\[4pt]&
-\frac{\pi}{16\,\delta\hti}\,\frac{1-(1+2\,\hk^2\,\delta\hti)\,\exp(-2\,\hk^2\,\delta\hti)}{\hk^7}\,
\notag\\[4pt]&\hskip5mm\times
\sum_{n_1,n_2,n_3=1}^{N-1}
\sum_{\ m_1,m_2,m_3=0}^{1}
 \GNodes(n_1+m_1,n_2+m_2,n_3+m_3,\hti)\,M_{m_1 m_2 m_3}(n_1,n_2,n_3;\hk)
\notag\\[4pt]&
+\frac{\pi}{16\,\delta\hti}\,\frac{1-2\,\hk^2\,\delta\hti-\exp(-2\,\hk^2\,\delta\hti)}{\hk^7}\,
\notag\\[4pt]&\hskip5mm\times
\sum_{n_1,n_2,n_3=1}^{N-1}
\sum_{\ m_1,m_2,m_3=0}^{1}
 \GNodes(n_1+m_1,n_2+m_2,n_3+m_3,\hti+\delta\hti)\,M_{m_1 m_2 m_3}(n_1,n_2,n_3;\hk)
\label{twkkEvolutionRd_DiscretizedA}
\end{align}
Here, each $M_{m_1 m_2 m_3}(n_1,n_2,n_3;\hk)$ involves
  corresponding 1-dimensional integrals
    $\int_{\hk_{n_2}}^{\hk_{n_2+1}} d\hl$ and  $\int_{\hk_{n_3}}^{\hk_{n_3+1}} d\hl$
or 2-dimensional integrals (if the above-mentioned analytical integration not implemented)
$$\int_{\hk_{n_2}}^{\hk_{n_2+1}} d\hl\int_{\max\{|\hl-\hk|, \hk_{n_1}\}}^{\min\{1, \hl+\hk, \hk_{n_1+1}\}}d\hm,
\qquad
  \int_{\hk_{n_3}}^{\hk_{n_3+1}} d\hl\int_{\max\{|\hl-\hk|, \hk_{n_1}\}}^{\min\{1, \hl+\hk, \hk_{n_1+1}\}}d\hm
$$
The details of $M_{m_1 m_2 m_3}(n_1,n_2,n_3;\hk)$ are not given here due to the cumbersomeness
and they can be easily inferred from the procedure listed above.

We can now select a finite set of collocation points for $\hk\in(0,1)$ to enforce 
\eqref{twkk_NonNegative_Submodel},
\begin{align}
 \tWAsy_{kk}(\hk_m,\hti+\delta\hti)\geq 0,\ \ m=1,\cdots,M
\end{align}
%
%
For the evaluation of \eqref{wwIneq_Submodel} through \eqref{vvIneq_Submodel}, 
we resort to the straight-forwardly derived
\begin{align}
&
\frac{1}{2\pi}\,\WAsy_{ij}(\hbr,\hti+\delta\hti)
\notag\\[4pt]
=\,&
 \int_0^{1}d\hk\,F_{ij}(\hbr,\hk)\,
     \exp(-2\,\hk^2\,\delta\hti)\,\tWAsy_{kk}(\hk,\hti)
\notag\\[4pt]&
-\frac{\pi}{16\,\delta\hti}
\sum_{n_1,n_2,n_3=1}^{N-1}
\sum_{\ m_1,m_2,m_3=0}^{1}
 \GNodes(n_1+m_1,n_2+m_2,n_3+m_3,\hti)\,
\notag\\[4pt]&\hskip35mm\times
 \int_0^{1}d\hk\,F_{ij}(\hbr,\hk)\,
     \frac{1-(1+2\,\hk^2\,\delta\hti)\,\exp(-2\,\hk^2\,\delta\hti)}{\hk^7}\,
         M_{m_1 m_2 m_3}(n_1,n_2,n_3;\hk)
\notag\\[4pt]&
+\frac{\pi}{16\,\delta\hti}
\sum_{n_1,n_2,n_3=1}^{N-1}
\sum_{\ m_1,m_2,m_3=0}^{1}
 \GNodes(n_1+m_1,n_2+m_2,n_3+m_3,\hti+\delta\hti)\,
\notag\\[4pt]&\hskip35mm\times
 \int_0^{1}d\hk\,F_{ij}(\hbr,\hk)\,
    \frac{1-2\,\hk^2\,\delta\hti-\exp(-2\,\hk^2\,\delta\hti)}{\hk^7}\,
        M_{m_1 m_2 m_3}(n_1,n_2,n_3;\hk)
\label{RepresentationWij(br)_Discretized}
\end{align}
where
\begin{align}
F_{ij}(\hbr,\hk)
:=
 \delta_{ij}\, \hk \bigg(\frac{\sin(\hr \hk)}{\hr}
                         -\frac{\sin(\hr\hk)}{\hr^3 \hk^2}
                        +\frac{\cos(\hr\hk)}{\hr^2 \hk}\bigg)
-\frac{\hr_i \hr_j}{\hr^5}\,\frac{(\hr^2 \hk^2-3) \sin(\hr\hk)
                                   +3 \hr \hk \cos(\hr\hk)}{\hk}
\end{align}

With the help of \eqref{twkkEvolutionRd_DiscretizedA},
the objective function of \eqref{ObjectiveFunction_Submodel} can be recast, at $\hti+\delta\hti$,
in the equivalent form of
\begin{align}
\sum_{n_1,n_2,n_3=1}^{N-1}
\sum_{\ m_1,m_2,m_3=0}^{1}
\,&
 \GNodes(n_1+m_1,n_2+m_2,n_3+m_3,\hti+\delta\hti)\,
\notag\\[4pt]&\hskip5mm\times
    \int_{0}^{1} d\hk\,\frac{1-2\,\hk^2\,\delta\hti-\exp(-2\,\hk^2\,\delta\hti)}{\hk^5}\,M_{m_1 m_2 m_3}(n_1,n_2,n_3;\hk)
\label{ObjectiveFunction_DiscretizedA}
\end{align}

To carry out the numerical simulation of the discretized SOCP problem above,
 we need to start from $\hti=0$ and fix the time step $\delta\hti$.
We notice the inconsistency between this  $\hti=0$ and the supposed asymptotic states at large time;
Computationally, we adopt appropriate initial conditions for
 $\tWAsy_{kk}(\hk,0)$ and $\GNodes(n_1,n_2,n_3,0)$ satisfying the discretized constraints above,
 solve for $\GNodes(n_1,n_2,n_3,\delta\hti)$ and $\tWAsy_{kk}(\hk,\delta\hti)$ through the optimization,
  sequentially in time,
 and expect the asymptotic state solutions to emerge from these artificial transient states.

The dimensionless turbulent energy at $\hti+\delta\hti$ is 
\begin{align}
&
\WAsy_{kk}(\bo,\hti+\delta\hti)
%
\notag\\[4pt]
=\,&
 4\pi \int_{0}^{1} d\hk\,\hk^2\,\exp(-2\,\hk^2\,\delta\hti)\,\tWAsy_{kk}(\hk,\hti)
\notag\\[4pt]&
-\frac{\pi^2}{4\,\delta\hti}
  \sum_{n_1,n_2,n_3=1}^{N-1}\sum_{\ m_1,m_2,m_3=0}^{1}
        \GNodes(n_1+m_1,n_2+m_2,n_3+m_3,\hti)\,
\notag\\[4pt]&\hskip30mm\times
   \int_{0}^{1} d\hk\,\frac{1-(1+2\,\hk^2\,\delta\hti)\,\exp(-2\,\hk^2\,\delta\hti)}{\hk^5}\,
                 M_{m_1 m_2 m_3}(n_1,n_2,n_3;\hk)
\notag\\[4pt]&
+\frac{\pi^2}{4\,\delta\hti}
  \sum_{n_1,n_2,n_3=1}^{N-1}\sum_{\ m_1,m_2,m_3=0}^{1}
        \GNodes(n_1+m_1,n_2+m_2,n_3+m_3,\hti+\delta\hti)\,
\notag\\[4pt]&\hskip30mm\times
  \int_{0}^{1} d\hk\,\frac{1-2\,\hk^2\,\delta\hti-\exp(-2\,\hk^2\,\delta\hti)}{\hk^5}\,
                M_{m_1 m_2 m_3}(n_1,n_2,n_3;\hk)
\label{wkkEvolutionRd_DiscretizedA}
\end{align}
which can be computed with the known $\tWAsy_{kk}(\hk,\hti)$ and $\GNodes(n_1,n_2,n_3,\hti)$
 and the newly solved $\GNodes(n_1,n_2,n_3,\hti+\delta\hti)$.

\end{document}